\documentclass[aps,pre,twocolumn,superscriptaddress,showpacs]{revtex4-1}
\usepackage{amsmath}
\usepackage{amssymb}
\usepackage{tikz}
\usepackage{graphicx}
\usepackage{dcolumn}
\usepackage{bm}
\usepackage{epsfig}
\usepackage{psfrag}
\def\II{\hbox{$1\hskip -1.2pt\vrule depth 0pt height 1.6ex width 0.7pt\vrule depth 0pt height 0.3pt width 0.12em$}}
\newcommand{\reffig}[1]{\mbox{Fig.~\ref{#1}}}
\newcommand{\refeq}[1]{\mbox{Eq.~(\ref{#1})}}
\newcommand{\refsec}[1]{\mbox{Sec.~\ref{#1}}}

\newcommand{\be}{\begin{equation}}
\newcommand{\ee}{\end{equation}}
\newcommand{\bal}{\begin{align}}
\newcommand{\eal}{\end{align}}
\newcommand{\ba}{\begin{eqnarray}}
\newcommand{\ea}{\end{eqnarray}}
\renewcommand{\Re}{\mathrm{Re}}

\newcommand{\T}{${\mathcal T}\,$}
\newcommand{\Ti}{${\mathcal T}$}
\def\II{\hbox{$1\hskip -1.2pt\vrule depth 0pt height 1.6ex width 0.7pt\vrule depth 0pt height 0.3pt width 0.12em$}}

\begin{document}

\title{\bf Experimental study of closed and open microwave waveguide graphs with preserved and partially violated time-reversal invariance}
\author{Weihua Zhang}
\email{zhangwh18@lzu.edu.cn}
\address{%
Lanzhou Center for Theoretical Physics and the Gansu Provincial Key Laboratory of Theoretical Physics, Lanzhou University, Lanzhou, Gansu 730000, China
}
\address{%
Center for Theoretical Physics of Complex Systems, Institute for Basic Science (IBS), Daejeon 34126, Korea
}
\author{Xiaodong Zhang}
\address{%
Lanzhou Center for Theoretical Physics and the Gansu Provincial Key Laboratory of Theoretical Physics, Lanzhou University, Lanzhou, Gansu 730000, China
}
\author{Jiongning Che}
\address{%
Lanzhou Center for Theoretical Physics and the Gansu Provincial Key Laboratory of Theoretical Physics, Lanzhou University, Lanzhou, Gansu 730000, China
}
\author{Junjie Lu}
\address{%
Institut de Physique de Nice, CNRS UMR 7010, Universit\'e C${\hat o}$te d'Azur, 06108 Nice, France
}
\author{M. Miski-Oglu}
\address{%
GSI Helmholtzzentrum f\"ur Schwerionenforschung GmbH
D-64291 Darmstadt, Germany}
\author{Barbara Dietz}
\email{corr. author: dietz@lzu.edu.cn}
\address{%
Lanzhou Center for Theoretical Physics and the Gansu Provincial Key Laboratory of Theoretical Physics, Lanzhou University, Lanzhou, Gansu 730000, China
}
\address{%
Center for Theoretical Physics of Complex Systems, Institute for Basic Science (IBS), Daejeon 34126, Korea
}

\date{\today}

\bigskip

\begin{abstract}
We report on experiments that were performed with microwave waveguide systems and demonstrate that in the frequency range of a single transversal mode they may serve as a model for closed and open quantum graphs. These consist of bonds that are connected at vertices. On the bonds, they are governed by the one-dimensional Schr\"odinger equation with boundary conditions imposed at the vertices. The resulting transport properties through the vertices may be expressed in terms of a vertex scattering matrix. Quantum graphs with incommensurate bond lengths attracted interest within the field of quantum chaos because, depending on the characteristics of the vertex scattering matrix, its wave dynamic may exhibit features of a typical quantum system with chaotic counterpart. In distinction to microwave networks, which serve as an experimental model of quantum graphs with Neumann boundary conditions, the vertex scattering matrices associated with a waveguide system depend on the wavenumber and the wave functions can be determined experimentally. We analyze the spectral properties of microwave waveguide systems with preserved and partially violated time-reversal invariance, and the properties of the associated wave functions. Furthermore, we study properties of the scattering matrix describing the measurement process within the frame work of random matrix theory for quantum chaotic scattering systems. 
\end{abstract}

\bigskip
\maketitle

\section{Introduction\label{Intro}} 

Quantum graphs~\cite{Kottos1997,Kottos1999} have served for three decades as a suitable system for the study of features of quantum systems, whose corresponding classical dynamic is fully chaotic. Linus Pauling introduced them for the modeling of organic molecules \cite{Pauling1936} and they are also employed to simulate  a large variety of other physical systems like, e.g. quantum wires~\cite{Sanchez1998,Kostrykin1999}, optical waveguides~\cite{Mittra1971} and mesoscopic quantum systems~\cite{Kowal1990,Imry1996}. They are constructed from bonds that are connected at vertices~\cite{Kottos1997,Kottos1999,Pakonski2001,Texier2001,Kuchment2004,Gnutzmann2006,Berkolaiko2013}. Wave propagation in a quantum graph is governed by the one-dimensional Schr\"odinger equation along the bonds with boundary conditions at their ends, that is, at the vertices, that ensure continuity of the wave functions and current conservation. It has been proven rigorously in Ref.~\cite{Gnutzmann2004} that, depending on the boundary conditions, closed quantum graphs with incommensurate bond lengths exhibit in their eigenvalue spectra the fluctuation properties of typical quantum systems with chaotic classical dynamic~\cite{Bohigas1984,Guhr1998,Haake2018,Heusler2007}. According to the Bohigas-Gianonni-Schmit conjecture (BGS)~\cite{Bohigas1984,Guhr1998,Haake2018,Heusler2007} these are described by the Gaussian ensembles of random matrix theory (RMT)~\cite{Mehta1990}. The boundary conditions can be expressed in terms of unitary vertex matrices~\cite{Kottos1999,Kostrykin1999,Gnutzmann2006,Harrison2007,Lawniczak2019}, that characterize the transport or scattering properties of the waves through the vertices. It was shown based on an exact trace formula~\cite{Keating1991,Kottos1999,Roth1983} that ergodicity of the wave dynamic results from the scattering characteristics of the waves entering and exiting a vertex through the bonds connected to it~\cite{Gnutzmann2006}. Also the two-point correlation functions of the scattering matrix associated with the scattering dynamic of open graphs that are coupled to their environment through leads, i.e., bonds that extend to infinity, were shown to coincide with those of random matrices applicable to typical quantum-chaotic scattering systems~\cite{Verbaarschot1985,Pluhar2013,Pluhar2013a,Pluhar2014,Fyodorov2005}. Thus, even though closed and open quantum graphs are basically described by the one-dimensional Schr\"odinger equation,  their wave dynamic may exhibit a rich variety of features observed in quantum systems with a chaotic classical dynamic. Furthermore, they are mathematically simple in the sense, that a secular equation can be written down explicitly for their eigenstates~\cite{Kottos1999}, so that these can be determined numerically with much less efforts than is required, e.g., for quantum billiards~\cite{LesHouches1989,StoeckmannBuch2000,Haake2018,Texier2001}, which are also accessible experimentally~\cite{Sridhar1991,Graef1992,Stein1992,So1995,Deus1995,Stoeckmann2001,Dietz2015a}. 

Another advantage of quantum graphs is, that all three universality classes associated with Dyson's threefold way~\cite{Dyson1962} can be simulated experimentally for Neumann boundary conditions or, generally, $\delta$-type boundary conditions at the vertices~\cite{Kottos1997,Kottos1999,Pakonski2001,Kuchment2004}  with microwave networks~\cite{Hul2004}, which are composed of coaxial cables corresponding to the bonds that are coupled by joints at the vertices. Note that these are wave-dynamical systems, however, the BGS conjecture also applies to systems exhibiting wave chaos~\cite{StoeckmannBuch2000,Dembowski2002}. Experiments with microwave networks with preserved time-reversal (\Ti) invariance, which belong to the orthogonal universality class, that is, with an antiunitary symmetry $\boldsymbol{\hat T}$ with $\boldsymbol{\hat T}^2=1$, and with violated \T invariance, i.e., unitary universality class, revealed~\cite{Hul2004,Lawniczak2010,Bialous2016} that, indeed, the fluctuation properties in their spectra agree well with those of random matrices from the Gaussian orthogonal ensemble (GOE) and the Gaussian unitary ensemble (GUE), respectively. Above all, microwave networks can be employed to model experimentally quantum systems with an antiunitary symmetry with $\boldsymbol{\hat T}^2=-1$~\cite{Rehemanjiang2016,Martinez2018,Martinez2019,Lu2020,Che2021}, whose spectral fluctuations coincide with those of random matrices from the Gaussian symplectic ensemble (GSE)~\cite{Scharf1988,Haake2018}. Only recently, the universality classes of microwave-network realizations~\cite{Rehemanjiang2020} could be extended to the ten-fold way~\cite{Altland1997}. The properties of open quantum graphs with wave chaotic dynamic have been investigated experimentally in ~\cite{Lawniczak2008,Lawniczak2011,Hul2012,Lawniczak2014,Hul2012,Lawniczak2020,Chen2021}. 

A drawback of microwave networks and quantum graphs with Neumann boundary conditions is the presence of backscattering at their vertices that leads to eigenstates that are localized on single bonds or on loops formed by a fraction of the bonds. These do not exhibit the complexity required to achieve agreement with RMT predictions for typical quantum systems with chaotic classical counterpart and they are non-universal because they depend on the lengths of the bonds they are confined to. They can be prevented by an appropriate choice of the boundary conditions. This was one of the motivations for designing quantum waveguide systems as a model of quantum graphs. They consist of straight waveguides with Dirichlet boundary conditions at the walls, that are connected at junctions~\cite{Post2012,Exner2015,Gnutzmann2022}. In the frequency range of a single transversal mode the associated Schr\"odinger equation is one-dimensional along the bonds. Furthermore, in distinction to microwave networks and Neumann quantum graphs, the vertex scattering matrices describing the transport of the waves through the junctions depends on the wavenumber. 

We simulate such systems experimentally with flat, metallic microwave waveguides, also referred to as waveguide graphs in the sequel. Here, we exploit the analogy of the associated Helmholtz equation with the Schr\"odinger equation of the quantum waveguide system for microwave frequencies below a maximum frequency which is inversely proportional to the height of the waveguides~\cite{Sridhar1991,Graef1992,Stein1992,So1995,Deus1995,Stoeckmann2001}. Actually, in 2015 experiments were performed with superconducting waveguide graphs in the quantum chaos group of Achim Richter and BD and completed just before the laboratory was closed. They have been presented in various presentations, however, a publication is in preparation since then due to various incidents that led to delays. The manuscript will be submitted soon~\cite{Dietz2022}. In these experiments the eigenvalues of the corresponding quantum waveguide graph could be determined with high accuracy and also properties of the scattering matrix describing the measurement process, which is directly related to that of the corresponding open quantum graph. The waveguide system was designed such that the $k$-dependence of the vertex scattering matrix and backscattering in the junctions is minimized, yielding a relative angle of 120$^\circ$, and thus vertex valency three for planar waveguide systems. The lengths of the waveguide graphs are incommensurate. 

An advantage of the microwave waveguide systems used in the present paper with respect to superconducting ones and to microwave networks is that the wave-function intensities are experimentally accessible. We analyze the spectral properties and fluctuation properties of the scattering matrix of closed and open waveguide graphs with preserved and partially violated time-reversal (\Ti) invariance, and perform an in-depth study of the properties of the wave functions for the case of preserved \T invariance. Furthermore, we investigate the spectral properties in a frequency range where single and double transversal modes exist. Here, the analogy to a conventional quantum graph is lost~\cite{Gnutzmann2022}. We would like to mention that, recently, photonic-crystal graphs where proposed as another model for quantum graphs and studied numerically with COMSOL Multiphysics~\cite{Ma2021}. In such a system the metal walls are replaced by an arrangement of rods on a certain lattice structure, and the waves simulating the quantum graph are confined to a band gap, which occurs depending on the structure and defects introduced to realize the waveguide system.   

The paper is organized as follows. In \refsec{MWN} we introduce waveguide graphs, their experimental realization and the procedures that are used to determine wave functions and to induce \Ti-invariance violation. Then, in~\refsec{TI} we present our experimental results on the spectral properties of waveguide graphs in the regions of a single transversal mode and where single and two transversal modes coexist, and we review our results on statistical properties of the electric field intensity. Furthermore, we followed Refs.~\cite{Kaplan2001,Hul2009} to analyze further statistical measures for the wave function properties of quantum graphs, e.g., in terms of inverse participation ratios. In~\refsec{TIV} we investigate spectral properties for the case of partial \Ti-invariance violation. Finally, in~\refsec{Smatrix} we summarize the results on the properties of the scattering matrix associated with the measurement process which is employed to obtain resonance spectra. In~\refsec{Concl} we summarize our findings and discuss them.   

\section{The Microwave Waveguide Network\label{MWN}} 
\begin{figure}[!th]
\includegraphics[width=\linewidth]{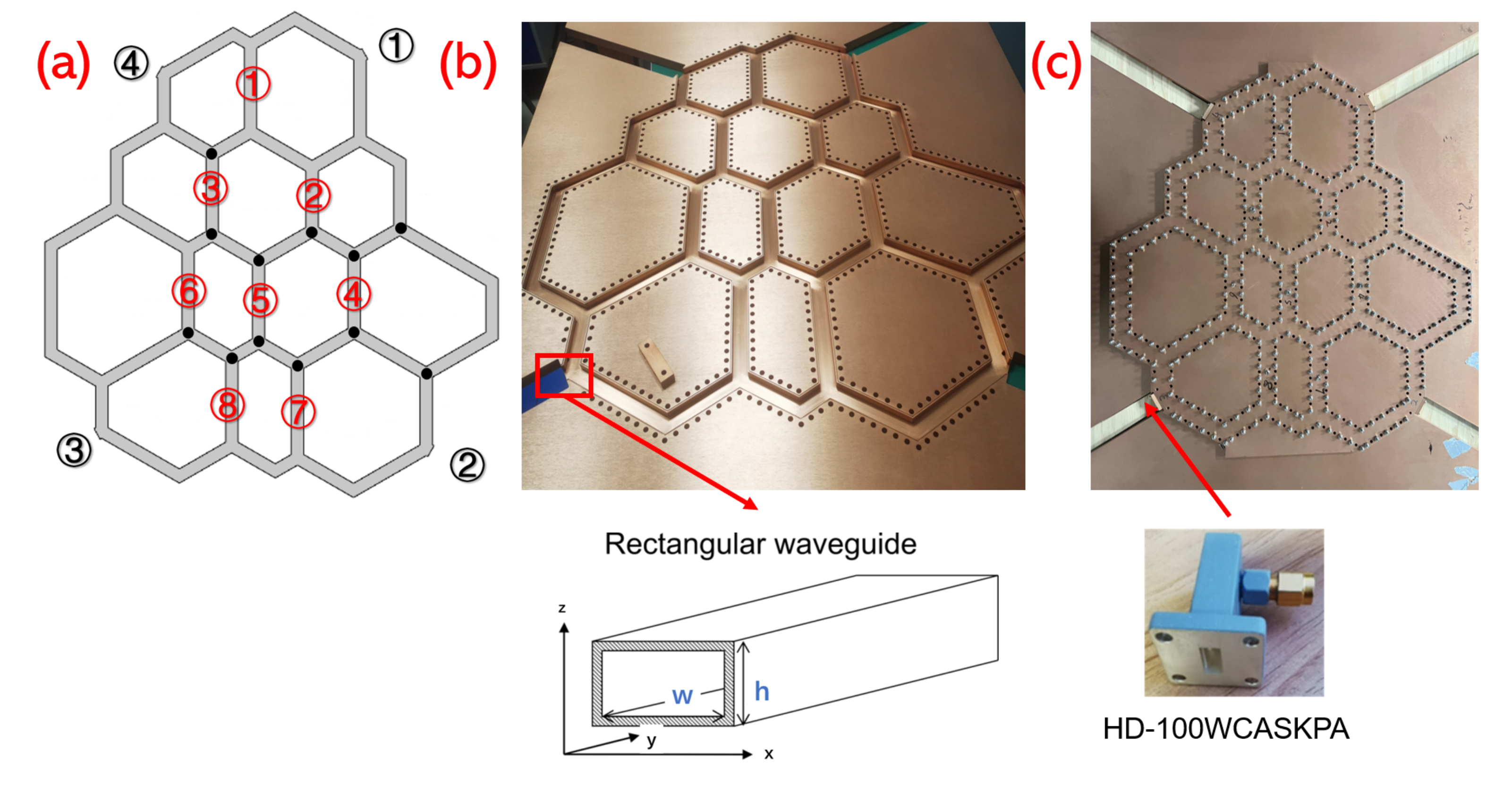}  
\caption{(a) Schematic view of the waveguide graph used in the experiment. Red numbers indicate the positions of the antennas, black ones those of the waveguide-to-coaxial adapters. The black disks are at the positions of the ferrites that are inserted into the waveguides and magnetized with exterior magnets to induce \Ti-invariance violation. (b) Bottom plate of the waveguide graph, exhibiting the channel system, the screw holes and the four cutouts, where the waveguide-to-coaxial adapters are attached. The lower part shows a sketch of the waveguides constructed by closing the channels with a plate. Both plates are manufactured from aviation aluminium and coated with a copper cover to attain higher quality factors. (c) Top view of the waveguide graph. The lower part shows a photograph of a waveguide-to-coaxial adapter.}   
\label{fig:sketch}    
\end{figure}

\begin{figure}[!th]
\includegraphics[width=1.0\linewidth]{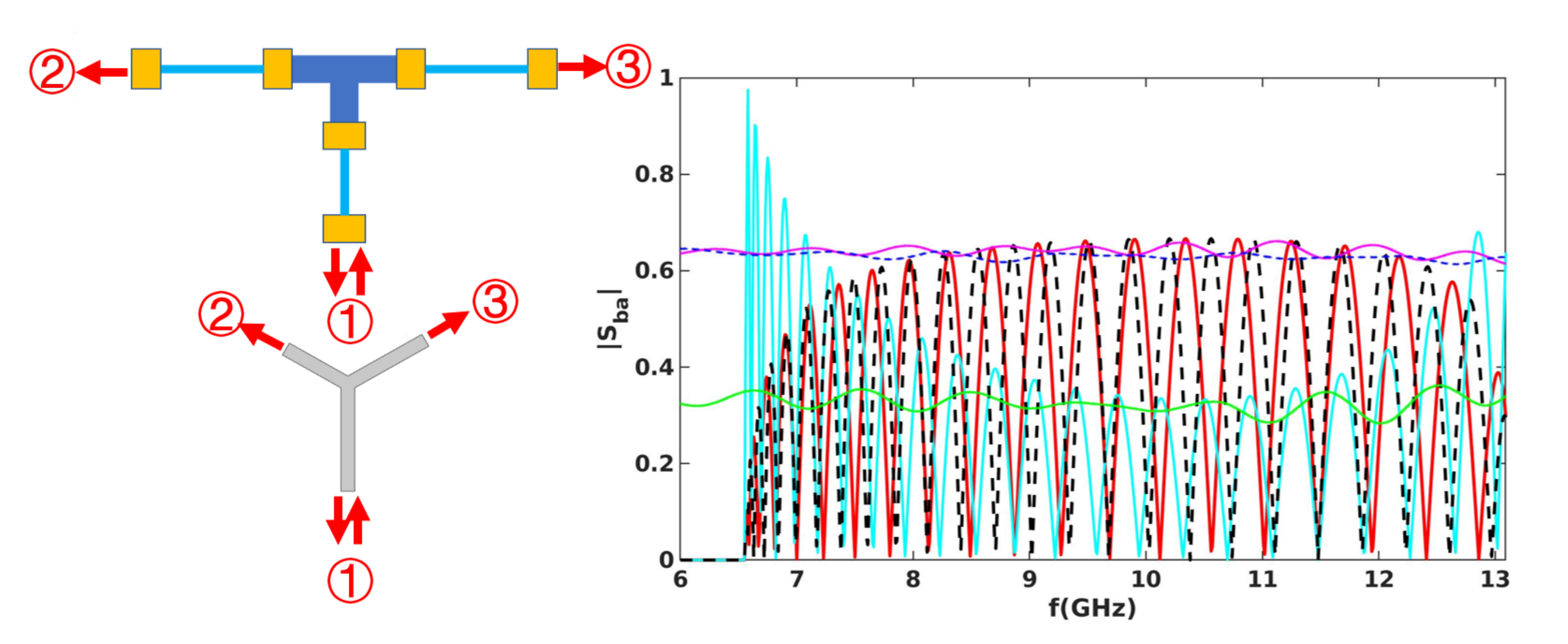}
\caption{Computed transmission and reflection spectra (right part) of a waveguide graph consisting of three waveguides of incommensurate lengths as shown schematically in the left part. Shown are the reflection spectra ($a=b=1$, cyan lines) and transmission spectra ($a=1,b=2$, black dashed lines) and  ($a=1,b=3$, red lines). The pink, dashed blue and green lines show the correponding measured transmission and reflection spectra, respectively, of a microwave network consisting of three coaxial cables that are connected by a conventional T joints. The spectra are shown in the frequency range, where only TM$_{10}$ modes exist, that is, where wave propagation takes place in the $(x,y)$ plane and is essentially one-dimensional.}
\label{fig:spara_3}
\end{figure}

The bonds of the microwave waveguide network are constructed from metallic rectangular waveguides of width $w$, height $h<w$ and incommensurate lengths $l_b\gg w$, as illustrated schematically in~\reffig{fig:sketch} (a) and in the left lower part. Respectively three of them are connected at a relative angle of $120^\circ$ at the vertices of the network. Below the cutoff frequency $f_{max}=c/2h$ for the second mode in the vertical ($z$) direction, only transverse-magnetic modes ${\rm TM_{n,0}},\, n=0,1,\dots$ are excited and the electric field strength $\vec E =E_z\vec e_z$, is perpendicular to the top and bottom of the waveguides. In that frequency range the microwaves are governed by the two-dimensional Helmholtz equation for a perfect electric conductor, that is, with Dirichlet boundary conditions at the side walls,
\be
\left[\frac{\partial^2}{\partial x^2}+\frac{\partial^2}{\partial y^2}+k^2\right]E_z=0,\, (x,y)\in\Omega,\, E_z\vert_{x,y\in\partial\Omega}=0,
\ee
with $x$ in the transversal direction and $y$ in the longitudinal one. Here, $E_z$ is the electric field strength in $z$ direction, $k=\frac{2\pi f}{c}$ denotes the wave number and $c$ the speed of light in vacuum. The wavenumbers in longitudinal direction corresponding to the ordered eigenfrequencies $f_m$, $f_1\leq f_2\leq\dots$ are given as 
\be\label{kym}
k_{y,m}=\sqrt{\left(\frac{2\pi f_m}{c}\right)^2-\left(\frac{n\pi}{w}\right)^2}
\ee
where the index $n$ counts the number of modes excited in transversal direction. For the width and height of the waveguides chosen in the experiment, $w=22.86$~mm,  $h=10.16$~mm the cutoff frequencies of the first and second transversal mode and for the first excited transverse-magnetic mode, ${\rm TM_{01}}$, are given as
\ba\label{Eq:fre}
	f_{\rm TM_{10}} &= \frac{c}{2w} &= 6.56\, {\rm GHz}, \\
	f_{\rm TM_{20}} &= \frac{c}{w} &= 13.12\, {\rm GHz}, \\
	f_{\rm TM_{01}} &= \frac{c}{2h} &= 14.76\, {\rm GHz},
\ea
respectively.
In the frequency range $f_{\rm TM_{10}}\leq f\leq f_{\rm TM_{20}}$ of single transversal modes the waveguide network simulates a quantum graph constructed from vertices with valency two at its bents and valency three at its junctions. The vertex scattering matrices associated with the boundary conditions at the vertices~\cite{Kottos1999,Kuchment2004,Gnutzmann2006,Berkolaiko2013} are obtained from the wave-function properties of the waveguide graph at the junctions. They depend strongly on $k$ and on the bending angle and were designed such that back scattering is minimized~\cite{Bittner2013}. This is in contrast to the vertex scattering matrix in the quantum graphs considered in~\cite{Kottos1999} and realized in the experiments with microwave networks~\cite{Hul2004}, where it corresponds to Neumann boundary conditions and thus is $k$-independent. In~\reffig{fig:spara_3} we show transmission (black dashed and red solid line) and reflection spectra (cyan line) that were computed with COMSOL Multiphysics for a waveguide graph consisting of three waveguides of incommensurate lengths, that are joined at a $120^\circ$ angle, as illustrated schematically in the lower left part of~\reffig{fig:spara_3}. They are shown in the frequency range, where only TM$_{10}$ modes exist, that is, where wave propagation takes place in the $(x,y)$ plane and is essentially one-dimensional. The period of the oscillations depends on the lengths of the waveguides and on the frequency $f$. They result from the superposition of waves of incommensurate periods entering the vertex. The waveguide graph is constructed from such subgraphs and  the wave chaotic features revealed in their spectral properties and in the fluctuation properties of the scattering matrix are attributed to this multi-connectivity yielding ergodicity of the microwave phases. For comparison we also show the measured transmission and reflection spectra of a microwave network consisting of three coaxial cables of the same geometric lengths as the waveguides that are joined by a conventional T joint~\cite{Martinez2018}, shown schematically in the upper left part of~\reffig{fig:spara_3}, which is characterized by a constant scattering matrix, $S_{ab}=2/3-\delta_{ab}$~\cite{Kottos1999,Hul2004}. For this case the oscillations are much less pronounced. Note, that the lengths of the waveguide graph correspond to the optical lengths of the coaxial cables, which are filled with a dielectric medium. 

The waveguide network comprising 48 waveguides with total length $\mathcal{L}\approx 5.814$~m along its central line, is constructed from a top ($934\times 868\times 17$~mm$^3$) and a bottom ($1200\times 1000\times 17$~mm$^3$) aviation aluminium plate. For the realization of the waveguides, channels of height $h$ and width $w$ are milled out of the bottom plate. A photograph is shown in~\reffig{fig:sketch} (b). A good electrical contact between the top and bottom plates is attained by screwing them tightly together through holes at distances 13~mm along the channels as recognizable in~\reffig{fig:sketch} (b) and (c). Furthermore, lead wire was inserted into grooves with width 1.3~mm and depth 1~mm that were milled out of the bottom plate along the channels. 

Resonance spectra of the waveguide network were measured with two procedures. For the first one eight wire antennas were attached to the top plate at the positions marked by red numbers in~\reffig{fig:sketch} (a). They are positioned at a distance of 1~mm from the central line. For the second procedure the waveguide graph was opened at two of the four bents, marked by black numbers, and waveguide-to-coaxial adapters (Model HD-100WCASKPA from HengDa MicroWave) were attached~\cite{Bittner2013}, referred to as ports in the sequel. The width $w$ and height $h$ of the waveguides are, actually, dictated by the impedance-matching condition with the adapters to ensure a reflectionless escape of microwaves through the ports. For the measurements, the antennas or the ports were connected to an Agilent N5227A vector network analyzer (VNA) via SUCOFLEX126EA/11PC35/1PC35 coaxial-cables sending microwaves into the resonator via one antenna $a$ or port and receiving it at the same or the other one, $b$. The VNA measures the relative phases $\phi_{ba}$ and ratios of the microwave power of the outcoming and ingoing rf signal, $\frac{P_{out,b}}{P_{in,a}}=|S_{ba}|^2$. Thereby, the complex scattering matrix element $S_{ba}=|S_{ba}|e^{i\phi_{ba}}$ describing the scattering process from antenna $a$ to antenna $b$ through the waveguide graph is obtained. 

It has been shown in~\cite{Albeverio1996} that the scattering matrix of a resonator coupled to a measuring apparatus, which in our case is the VNA, via $M$ leads supporting one open channel each, is given by 
\begin{equation}\label{eq:SResonator}
	\hat S(f) = \II - i\hat{W}^\dagger\left(f\II-\hat{H}^{Res}+\frac{i}{2}\hat W\hat W^\dagger\right)^{-1}\hat{W}.
\end{equation}
Here, $\hat H^{Res}$ denotes the Hamiltonian describing the closed resonator and $\hat W$ accounts for the coupling of the resonator modes to the $M$ open channels. In the vicinity of an isolated or weakly overlapping resonance at eigenfrequency $f_m$, $|S_{ba}|$ is well described by the complex Breit-Wigner form,
\begin{equation} \label{Eq:bw}
	S_{ba}(f)=\delta_{ba}-i\frac{\sqrt{\gamma_{ma}\gamma_{mb}}}{f-f_m+\frac{i}{2}\Gamma_m},
\end{equation}
where $\gamma_{na}$ and $\gamma_{nb}$ are the partial widths associated with antennas $a$ and $b$ and $\Gamma_n$ denotes the total width, which is given by the sum of the partial widths and the width $\Gamma_{abs}$ due to absorption in the walls of the waveguide.  The partial widths are proportional to the modulus of the wave functions at the positions of the antennas. The resonance parameters,  that is, the resonance strengths $\gamma_{na}\gamma_{nb}$, resonance widths $\Gamma_n$ and eigenfrequencies $f_n$ are determined by fitting the complex Breit-Wigner form~\refeq{Eq:bw} to the measured scattering matrix elements~\cite{Dembowski2005}. This is feasible, if the widths of the resonances are small compared to the average spacing between adjacent resonances. Consequently, a cavity with a high-quality factor $Q$ is a prerequisite. It depends on the absorption of microwave power in the walls, that is, the material, and it is proportional to the ratio of the volume to the surface of the resonator, that is, essentially to its height $h$ which is fixed by the size of the waveguide-to-coaxial adapters. To reduce absorption, both plates were coated with a copper cover of high conductivity, whose thickness 0.008~mm is much larger than the skin depth $\delta\approx 0.001$~mm. Thereby we attained quality factors of up to $Q\simeq 6500$, which was sufficient to identify complete sequences of eigenfrequencies in the relevant frequency regions. 

\begin{figure}[htbp]
\includegraphics[width=\linewidth]{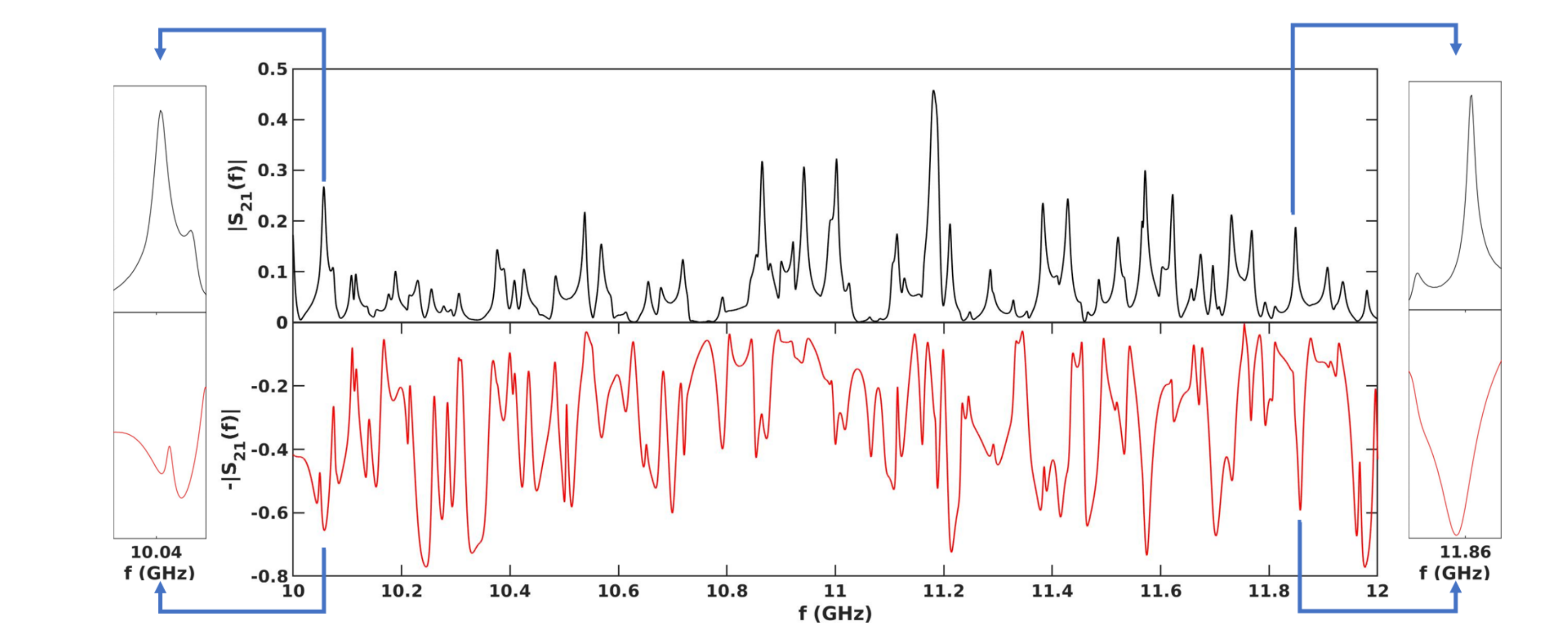}  
	\caption{Transmission spectra of the waveguide graph. The upper part shows the spectrum measured from antenna 1 to antenna 2 (black line), the lower one the spectrum measured from port 1 to port 2 (red line). The latter has been multiplied with (-1) in order to facilitate the comparison of the spectra. To the left and right are exhibited examples of weakly overlapping and isolated resonances, respectively.}   
\label{fig:s12}   
\end{figure}

Figure~\ref{fig:s12} shows parts of the transmission spectra measured from antenna 1 to antenna 2 (black line, top), and from port 1 to port 2 (red line, bottom), respectively. The latter was multiplied with (-1) to facilitate comparison of the spectra. Both spectra comprise isolated and weakly overlapping resonances, as illustrated to their right and left, respectively. The amplitudes are generally higher and a stronger resonance overlap is observed for the measurements with ports, which may be attributed to a larger opening of the waveguide system than in the measurements with antennas. Therefore, we used antennas for the determination of the eigenfrequencies.

As compared to microwave networks, waveguide graphs have the advantage that the wave function intensity distribution can not only be measured at the vertices but also along the waveguide parts. It is obtained from the electric field intensity distribution $E^2_z(x,y)$ which is determined  based on Slater’s theorem~\cite{Maier1952} by employing the perturbation body method~\cite{Sridhar1991,Doerr1998,Dembowski1999,Kuhl2007}. Namely, when introducing a metallic perturbation body into a microwave resonator a frequency shift is induced which depends on the squared electric and magnetic field at the position of the perturbation body,
\be\label{Eq:deltaf}
\Delta f(x,y)=f(x,y)-f_m=f_m\left[c_1 \vec E^2(x,y)-c_2\vec{B}^2(x,y)\right].
\ee
The constants $c_1$ and $c_2$ depend on the geometry and material of the perturbation body and $f_m$ denotes the resonance frequency of the resonator before introducing the perturbation body. The contribution of $\vec{B}(x, y)$ is removed by choosing a cylindrical perturbation body which is made from magnetic rubber (NdFeB)~\cite{Bogomolny2006}. It has a diameter of 8~mm and a height of 6~mm and is moved with an external magnet, which is fixed to a positioning unit which is described in~\cite{Zhang2019}, in steps of 3~mm parallel to the waveguide walls through the whole waveguide network. In order to determine the frequency shift $\Delta f(x,y)$ we determine at the eigenfrequency $f=f_m$ the difference $\Delta\phi$ of the relative phases between the received and emitted rf signal for the cases without and with perturbation body, which is proportional to $\Delta f$, so that $\Delta\phi\propto\Delta f\propto E_z^2(x,y)$~\cite{Dembowski1999}.

\begin{figure}[htbp]
	\includegraphics[scale=0.15]{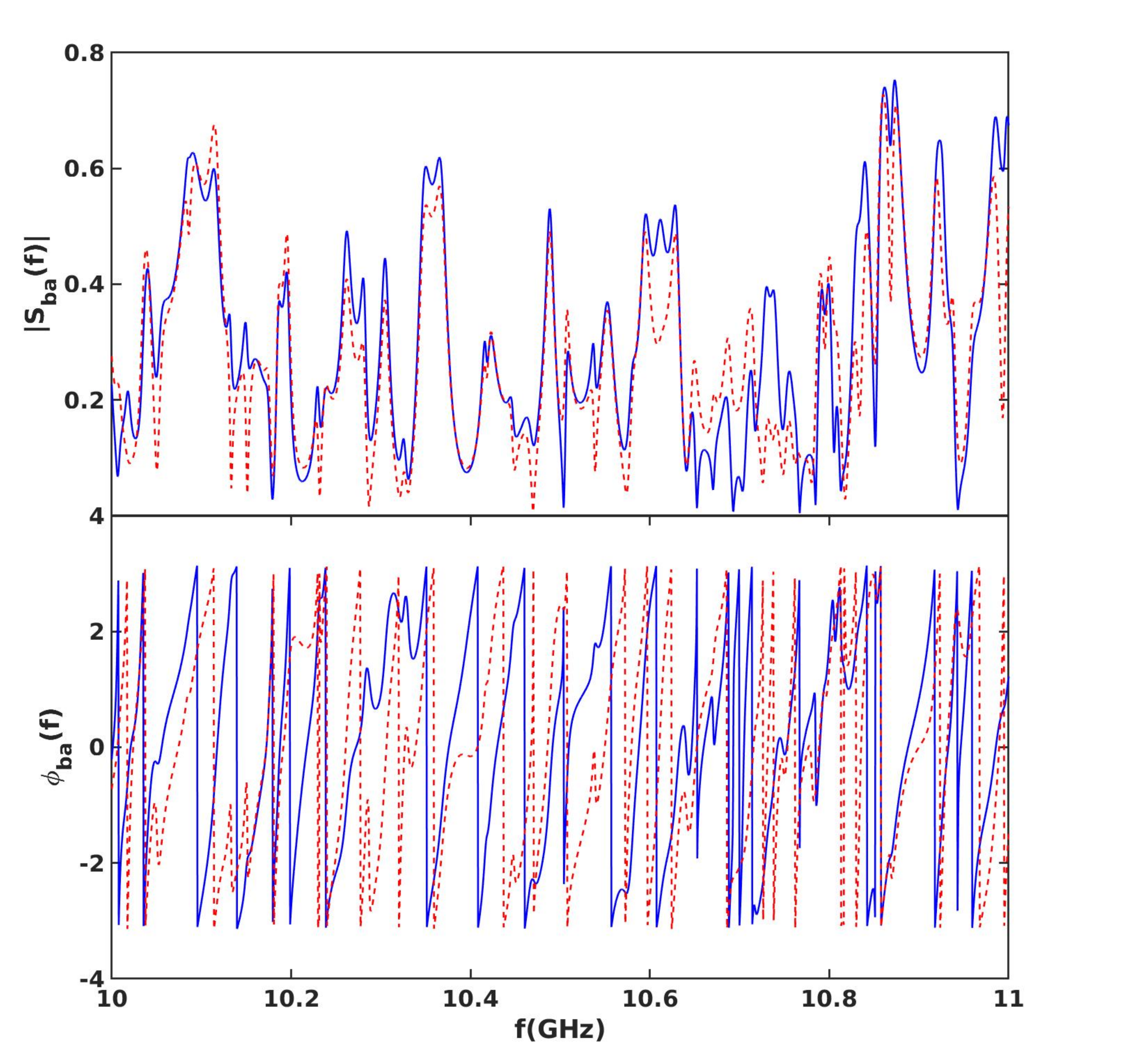} 
	\caption{Transmission spectra obtained in measurements with ports for the waveguide graph containing the magnetized ferrites. Shown are the amplitudes $|S_{ba}|$ (upper panel) and phases $\phi_{ba}$ (lower panel), respectively, for $a=1,\, b=2$ (red dashed lines) and $a=2,\, b=1$ (blue lines).}   
	\label{fig:ss12}   
\end{figure}

To induce violation of \T invariance in a microwave network, vertices are partly replaced by circulators~\cite{Hul2004,Lawniczak2010}, microwave devices with three ports which introduce a directionality in the sense that microwaves entering it at one port may only exit at one of the two other ports, respectively. Thereby, complete violation of \T invariance is induced. In the experiments with the waveguides we used the procedure of~\cite{So1995,Dietz2007,Dietz2009a}, that is, we inserted in total 12 19G3 cylinder-shaped ferrites made from Fe$_2$O$_3$ with diameter 5~mm and height 10~mm at the positions marked by black dots in~\reffig{fig:sketch} (a). Their saturation magnetization is $M_s=0.1941$~T. Each ferrite is magnetized by two external cylindrical NdFeB magnets of diameter 15~mm and height 20~mm, that are positioned above and below it to generate a uniform magnetic field in $z$ direction of strength $B\simeq 0.1264$~T. The magnetic field $B$ induces a macroscopic magnetization in the ferrites, thus causing a precession of the spins in the ferrite around it with the Larmor frequency. Violation of the principle of reciprocity~\cite{French1985,Ericson1966,Mahaux1966,Witsch1967,Blanke1983}, $S_{ab}(f)=S_{ba}(f)$, is induced through the coupling of the spins to the rf magnetic-field components of the resonator modes, whose size depends on the rotational direction of polarization of the latter. Since the modes are circularly polarized with unequal magnitudes of the two rotational components~\cite{Dietz2007}, this implies a deterioration of reciprocity between modes emitted at one antenna and received at another one and the reversed modes. Figure~\ref{fig:ss12} demonstrates that violation of the principles of detailed balance and reciprocity are attained when inserting the ferrites and magnetizing them. On the other hand, the principles hold for the waveguide graph without ferrites, since for that case the scattering matrix is symmetric. 
\section{Properties of the eigenvalues and wave functions of quantum waveguide graphs with preserved \T invariance\label{TI}}  
We use the analogy between quantum waveguide graphs~\cite{Post2012,Exner2015} and microwave waveguide networks of corresponding geometry to investigate their properties experimentally. In the experiments with antennas the transmission and reflection spectra were measured in a frequency range $f_{\rm TM_{10}}\leq f\leq f_{\rm TM{01}}$ in steps of 500~kHz, whereas the waveguide-to-coaxial converters operate in the frequency range $8.2\leq f\leq 12.4$~GHz and spectra were measured in steps of 400~kHz. The eigenfrequencies were determined by fitting the squared modulus of the Breit-Wigner form~\refeq{Eq:bw} to the resonances in the spectra $\vert S_{ba}(f)\vert^2$. For this their precise experimental determination is indispensable, that is, all systematic negative effects need to be removed. Dominant contributions to them come from the coaxial cables connecting the VNA with the cavity, which attenuate the rf signal and additional reflections occur at their interconnections with the VNA and resonator that complicate the extraction of the resonance parameters. These effects are removed by a proper calibration of the VNA before a measurement~\cite{Dembowski2005}. Furthermore, despite the coating with copper, there is absorption in the cavity walls, which leads to weakly overlapping resonances. The fitting procedure might fail in cases, where it is too strong or where two eigenfrequencies are lying too close to each other. Another cause for missing resonances are situations where the electric field strength is zero at the position of an antenna so that they cannot be excited. To avoid this, we performed measurements for various positions of the antennas. 

\begin{figure}[htbp]
\includegraphics[width=0.7\linewidth]{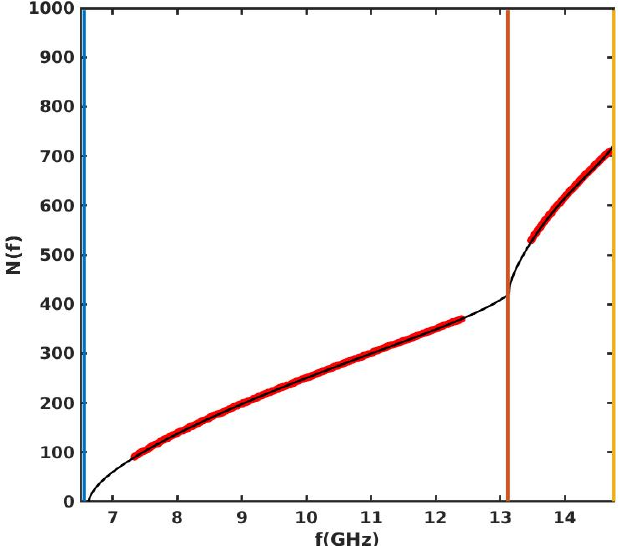}
	\caption{Integrated spectral density deduced from the experimental data (red circles) in comparison to the analytical results Eqs.~(\ref{Eq:n1}) and~(\ref{Eq:n2}) (black solid line). The red, blue and orange vertical lines indicate the locations of the cutoff frequencies $f_{\rm TM_{10}},\, f_{\rm TM_{20}},\, f_{\rm TM_{01}}$, respectively.}
\label{fig:Nf}
\end{figure}

In distinction to the experiments with microwave networks the spectral density, and thus the mean spacing depends on the eigenfrequency $f_m$, i.e., eigenwavenumber $k_m$. In the relevant frequeny ranges, the smooth part of the integrated spectral density is obtained from~\refeq{kym} as
\begin{equation}\label{Eq:n1}
N^{smooth}(k_m)=\frac{\mathcal{L}}{\pi}\sqrt{k_m^2-\left(\frac{\pi}{w}\right)^2}
\end{equation}
for $f_{\rm TM_{10}}\leq f\leq f_{\rm TM_{20}}$, and
\begin{equation}\label{Eq:n2}
	N^{smooth}(k_m)=\frac{\mathcal{L}}{\pi}\left[\sqrt{k_m^2-\left(\frac{2\pi}{w}\right)^2}+\sqrt{k_m^2-\left(\frac{\pi}{w}\right)^2}\right]
\end{equation}
for $f_{\rm TM_{20}}\leq f\leq f_{\rm TM_{01}}$. For frequencies much larger than the associated cutoff frequency $N^{smooth}(k)$ approaches that for the corresponding microwave network, $N^{smooth}(k)\simeq\frac{\mathcal{L}}{\pi}k$. In order to locate missing eigenfrequencies we looked at the difference of the number of identified eigenwavenumbers below $k_m$ and the expected number, $N(k_m)$. Missing levels manifest themselves as jumps in the locally averaged fluctuating part of the integrated spectral density $N(k)$, $N^{fluc}(k)=N(k)-N^{smooth}(k)$. We identified them and then carefully inspected all reflection and transmission spectra to check whether we oversaw a resonance because of the overlap with neighboring ones which would be visible as a bump in a resonance curve. In total 261 eigenfrequencies could be identified in the range 7.33-12.04~GHz for the antenna measurement, which coincides to the expected number within the range of error bars for the value of $\mathcal{L}$. We confirmed that we found all eigenfrequencies with simulations using COMSOL Multiphysics. In~\reffig{fig:Nf} we compare the analytical results Eqs.~(\ref{Eq:n1}) and~(\ref{Eq:n2}) (black line) to the smooth part of the experimentally obtained integrated spectral density (red circles). The curves agree very well, thus corroborating correctness of the unfolding procedure. 

\subsection{Fluctuation properties in the eigenfrequency spectrum for the single-mode case\label{SpProp}}

\begin{figure}[htbp]
\includegraphics[width=0.9\linewidth]{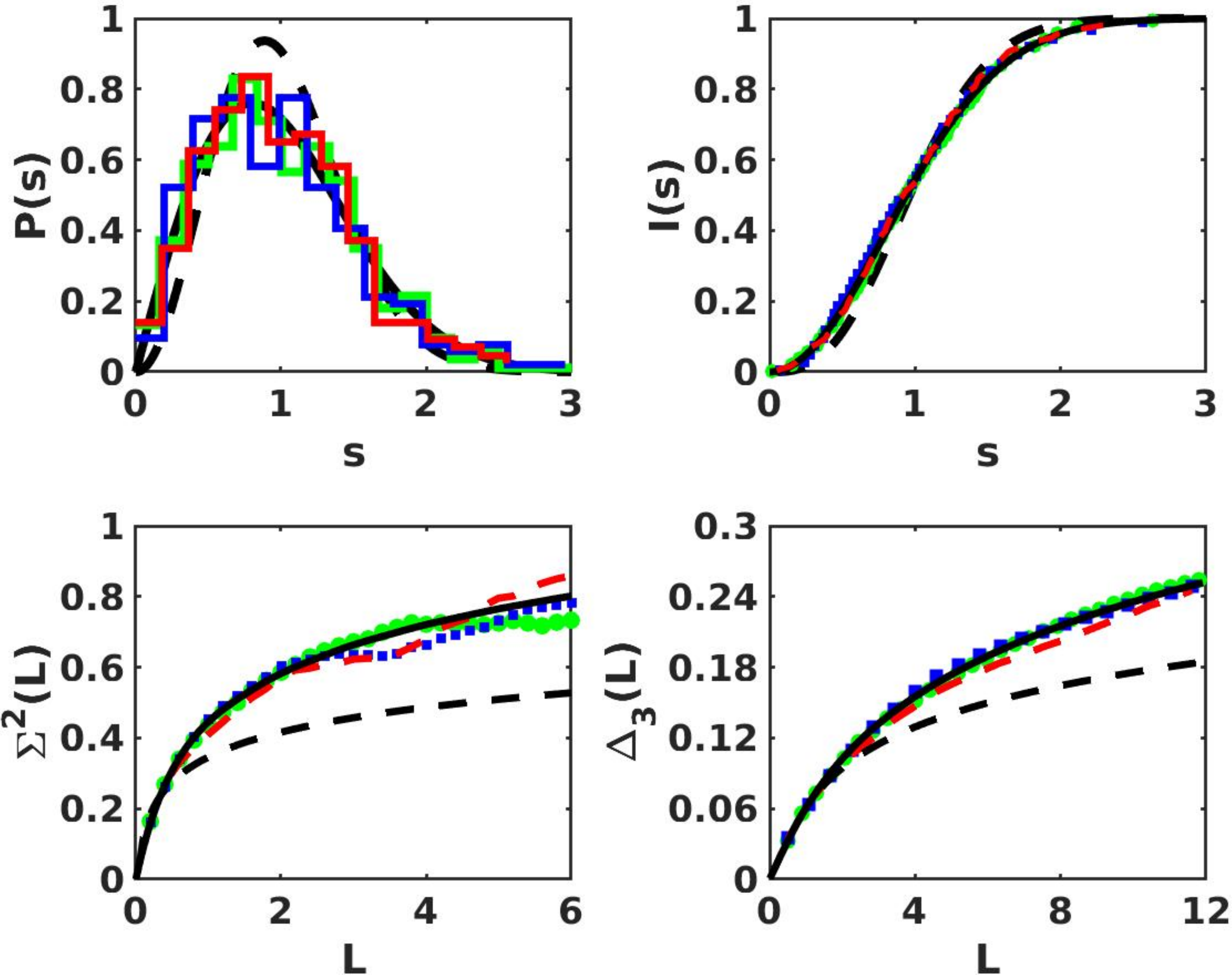} 
	\caption{Spectral Properties of the unfolded eigenfrequencies. Shown are the nearest-neighbor spacing distribution $P(s)$, the cumulative nearest-neighbor spacing distribution $I(s)$, the number variance $\Sigma^2(L)$ and spectral rigidity $\Delta_3(L)$. The red histogram and dashed lines, blue line and squares, and green lines and dots show the results deduced from the experimental data, COMSOL Multiphysics simulations and the quantum graph of corresponding geometry, respectively. The black full and dashed lines show the results for the GOE and GUE, respectively.}   
\label{fig:sps1d}   
\end{figure}

\begin{figure}[htbp]
	\includegraphics[width=0.9\linewidth]{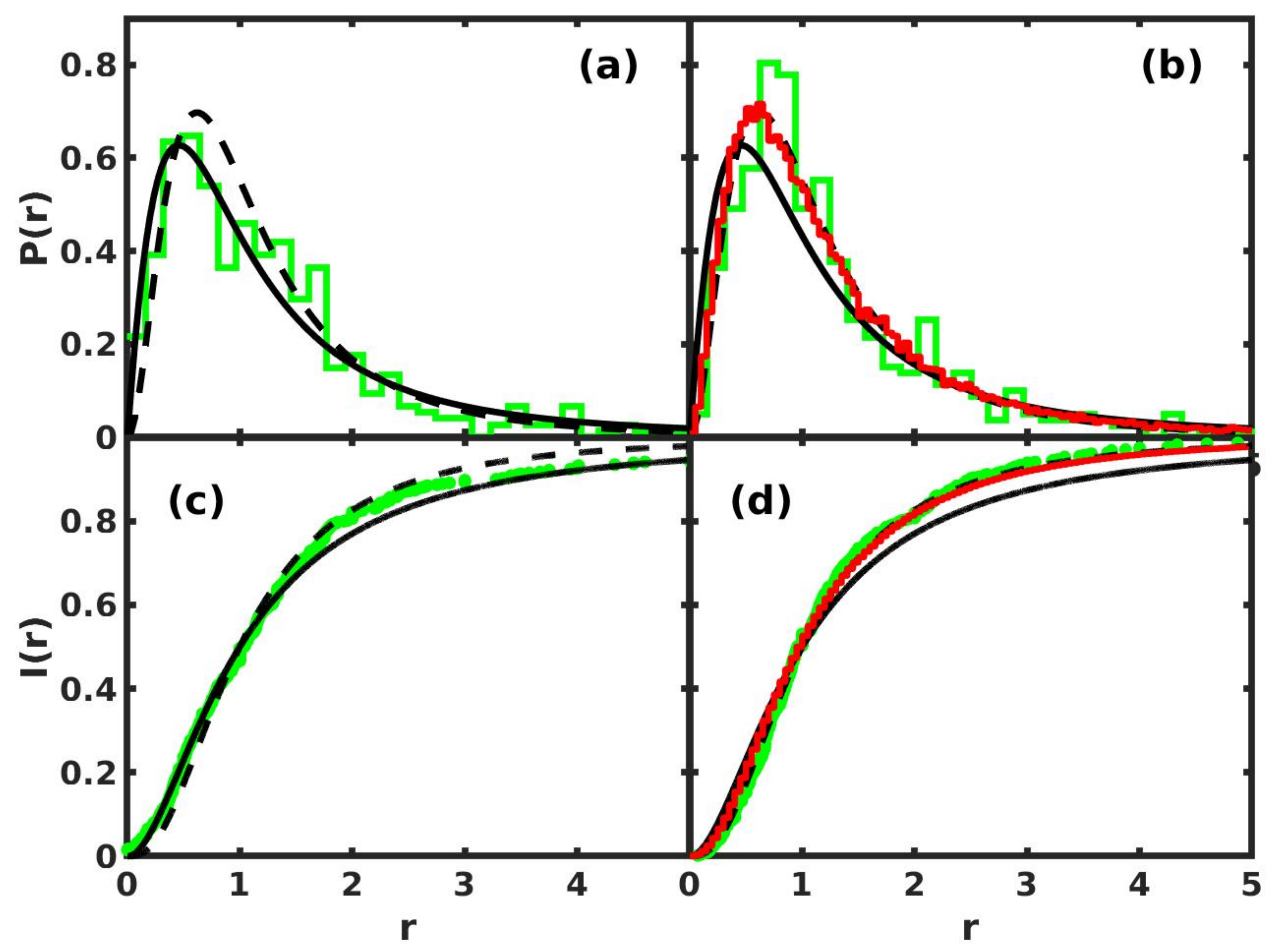}  
	\caption{Comparison of the ratio distributions (upper panels) and the cumulative ratio distributions (lower panels) of the experimentally (green histograms and dots) obtained results for the waveguide graph without [(a) and (c)] and with [(b) and (d)] partial violation of \T invariance, for the frequency range  8.7-14.5~GHz, where the \Ti-invariance violation parameter equals $\xi =0.3$. The corresponding RMT curves are shown as red lines. The black solid and dashed lines exhibit the results for random matrices from the GOE and GUE, respectively.}    
	\label{fig:pr}  
\end{figure}

The eigenwavenumbers $k_{y,m}$, obtained from the eigenfrequencies by employing~\refeq{kym}, were unfolded to mean spacing unity, that is, system specific properties were removed, by replacing them by $N^{smooth}(k_m)$ given in Eqs.~(\ref{Eq:n1}) and~(\ref{Eq:n2}). Furthermore, we computed 4500 eigenvalues for the corresponding quantum graph by proceeding as in~\cite{Kottos1999,Dietz2017} and those of the waveguide graph by employing COMSOL Multiphysics. Results for the spectral statistics are exhibited in~\reffig{fig:sps1d}. Shown are the distribution of the spacings between nearest-neighbor eigenvalues $P(s)$ and the associated cumulative distribution $I(s)$ as measures for short-range correlations, the number variance $\Sigma^2(L)$ and the Dyson-Mehta statistics $\Delta_3(L)$, which gives the spectral rigidity of a spectrum~\cite{Bohigas1975,Mehta1990}, as measures for long-range correlations.   
Agreement between the curves for GOE statistics and those of the quantum graph are good. For the experimental data (red curves) small deviations are observed for $P(s)$ and $I(s)$ which may be attributed to experimental inaccuracies. For $\Sigma^2(L)$ the agreement with GOE is similar to that for the simulations and for $\Delta_3(L)$ the curve lies below the GOE curve for $L\gtrsim 5$. A similar, but more pronounced, behavior has been observed for quantum graphs and microwave networks and has been attributed to the contributions from waves experiencing backscattering at vertices which leads to their confinement to individual bonds or a fraction of them~\cite{Dietz2017}. These are nonuniversal, as they depend on the lengths of the bonds and they do not exhibit the complexity of the dynamic which leads to the GOE like spectral properties. In the waveguides backscattering may result from reflections at the inner corners formed by the waveguides at the vertices~\cite{Bittner2013}. The distribution of the ratios of consecutive spacings of nearest-neighbor non-unfolded eigenfrequencies~\cite{Oganesyan2007,Atas2013,Atas2013a}, and the cumulative ratio distribution, plotted in~\reffig{fig:pr} (a) and (c), respectively, are also quite well described by the RMT results for the GOE.
Agreement of the spectral properties of the waveguide network with those of random matrices from the GOE indicates that they exhibit similar wave chaotic features as quantum graphs. In both cases the wave propagation is one dimensional along the bonds, so that the complexity is induced by the transport characteristics at the common junctions of the bonds. To further explore these features we investigated length spectra and measured the wave functions.   

\begin{figure}[htbp]
\includegraphics[width=\linewidth]{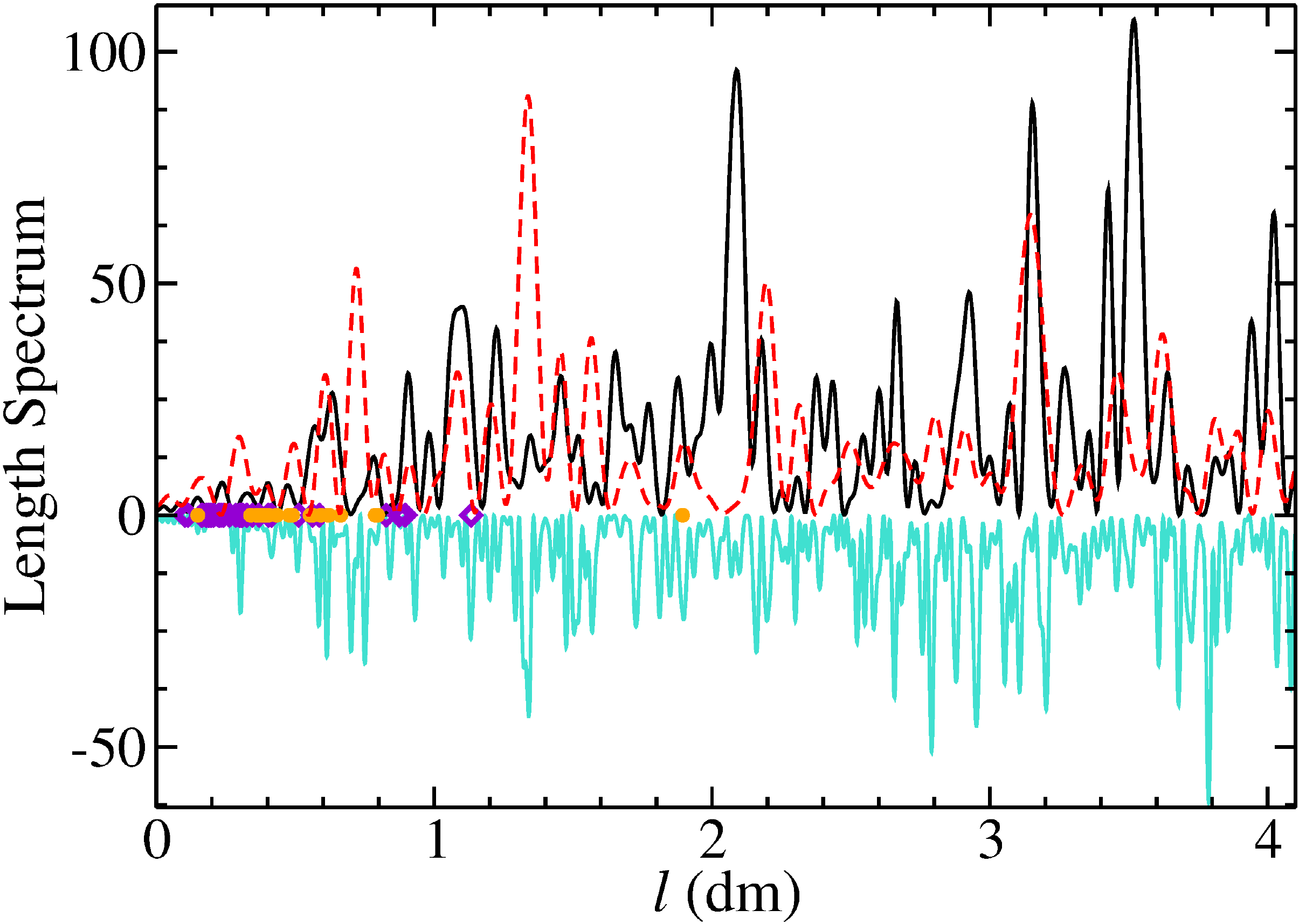}  
	\caption{Comparison of the length spectra deduced from the measurements with antennas (black line) with those obtained from the lowest 250 eigenvalues of the quantum graph of corresponding geometry (red line) and all computed eigenvalues (turquoise line). The violet diamonds mark twice the lengths of the bonds of the quantum graph and orange dots those of bonds connected to bents in the waveguide graph that lead to orbits not present in the corresponding quantum graph.}  
\label{fig:ls}  
\end{figure}

The occurrence of backscattering can be seen, e.g., in a length spectrum~\cite{Hul2004,Dietz2017}, which is obtained from the modulus of the Fourier transform of the fluctuating part of the spectral density from wavenumber to length and has the property that it exhibits peaks at the lengths of the periodic orbits of the corresponding classical system. In a quantum graph orbits are composed of itineraries along successive bonds that are uniquely defined by the sequence of the vertices connecting them. Similarly, in the waveguide graph the orbits correspond to the paths of the waves through the waveguide network. In~\reffig{fig:ls} we compare length spectra deduced from the eigenfrequencies which were determined from the antenna measurements (black solid lines) with those of the eigenvalues of the quantum graph of corresponding geometry taking into account a similar number of eigenvalues (red dashed lines) and also for all computed eigenvalues (turquoise line) where the trace formula provides a very good approximation of the spectral density. The length spectra of the quantum graph and waveguide graph differ in amplitude because of the distinct features of the vertex scattering matrices defining the wave transport through the vertices. Furthermore, in distinction to quantum graphs, the bents connecting two waveguides correspond to vertices in the waveguide networks, leading to additional peaks in their length spectra. The violet diamonds mark twice the lengths of the bonds in the quantum graph, orange dots twice the lengths of bonds connected to bents in the waveguide graph corresponding to these additional closed loops. Both length spectra exhibit peaks at lengths corresponding to twice the length of the bonds, thus indicating that backscattering is present.  

\subsection{Properties of the wave functions in the frequency range for the single-mode case\label{WFProp}}
\begin{figure}[htbp]
\includegraphics[width=0.4\linewidth]{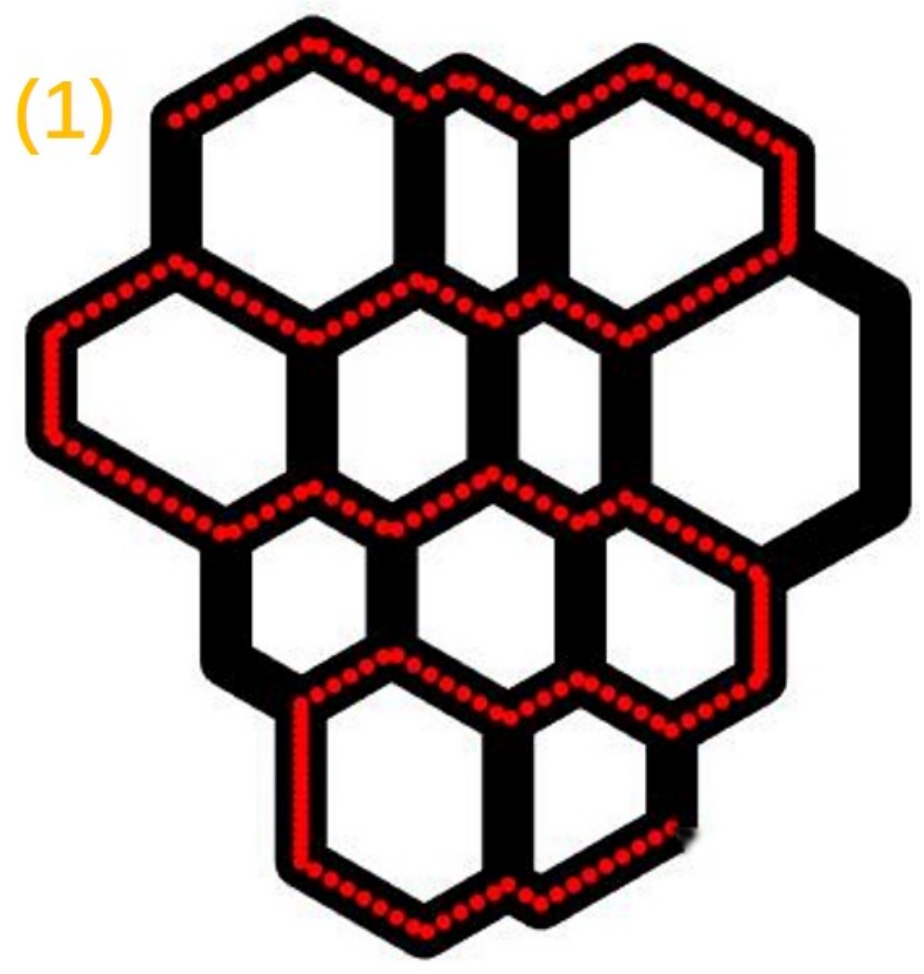}
\includegraphics[width=0.4\linewidth]{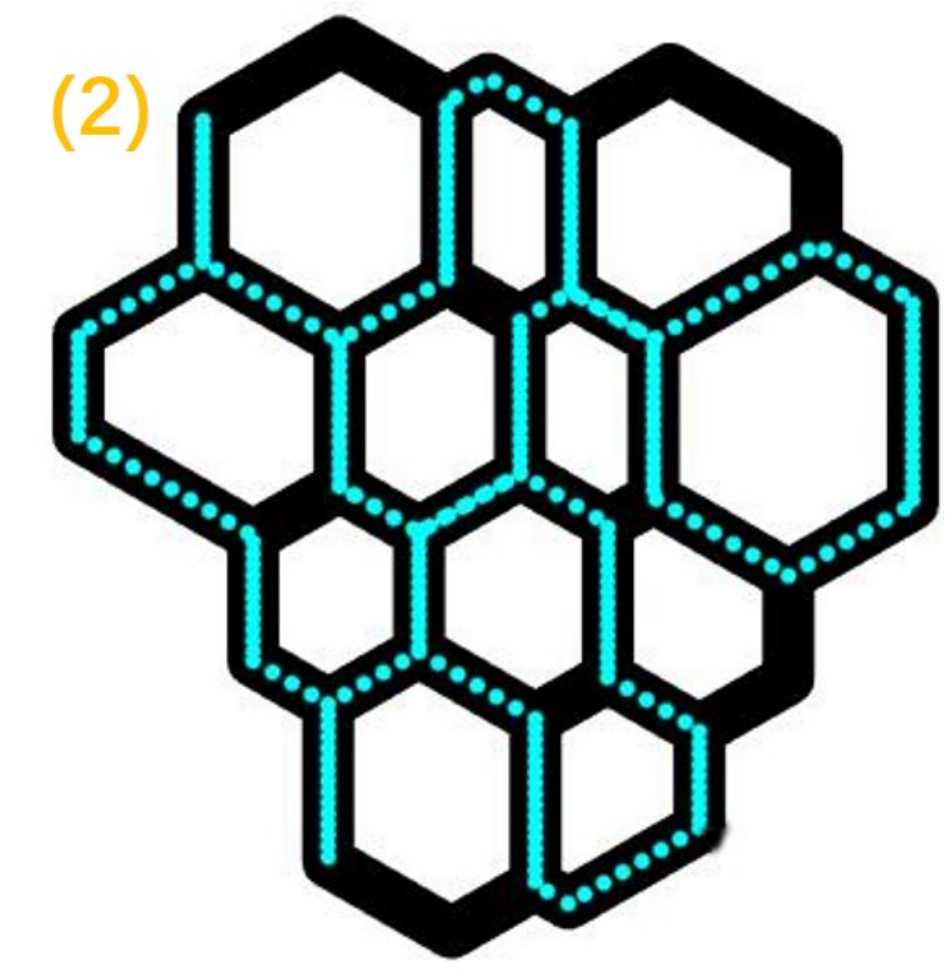}
\includegraphics[width=0.4\linewidth]{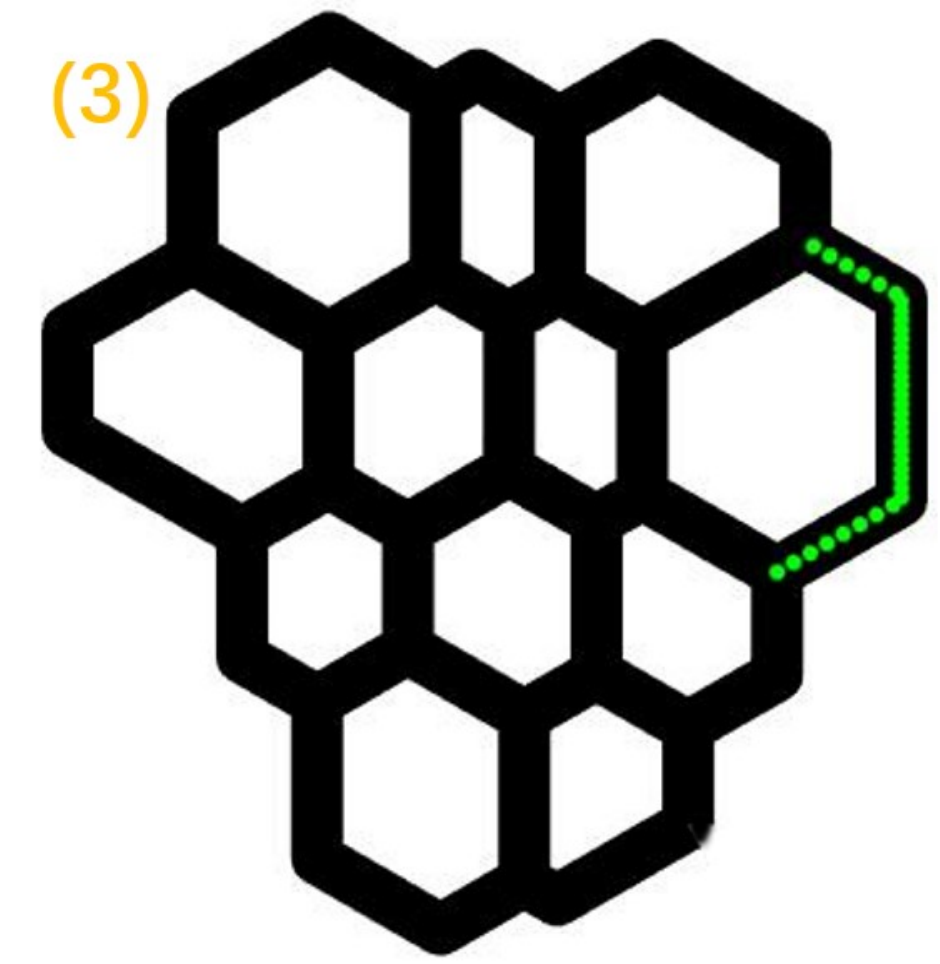}
\includegraphics[width=0.4\linewidth]{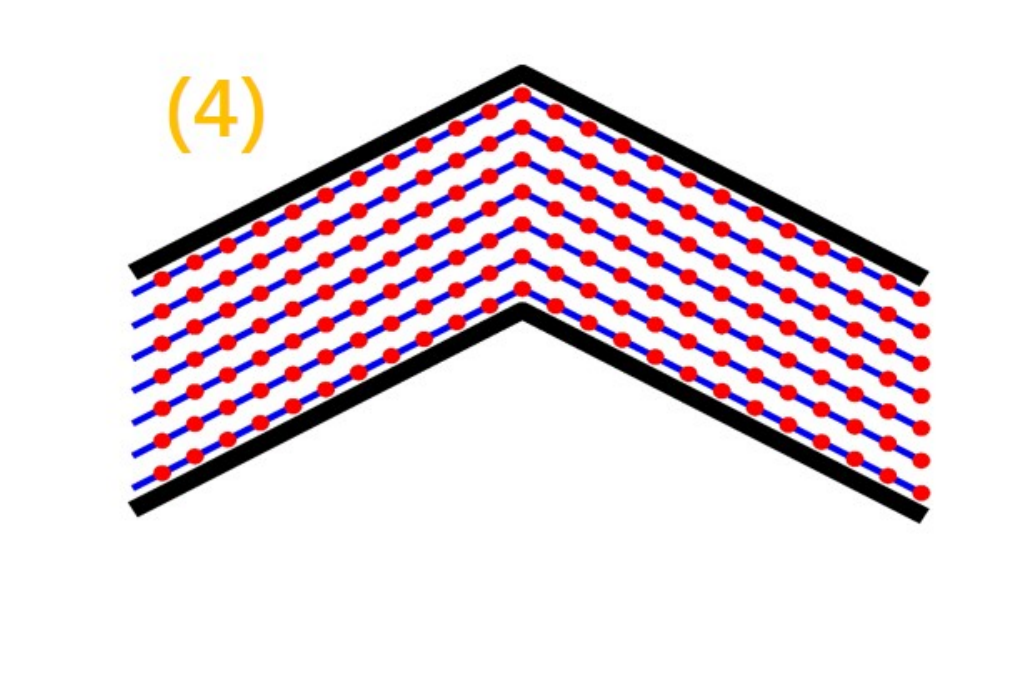}
	\caption{(1)-(3): Schematic view of the loops along which the perturbation body was guided. (4): Schematic view of the paths of the perturbation body which was moved along seven lines parallel to the waveguide walls.}
\label{Loops}
\end{figure}

\begin{figure}[htbp]
\includegraphics[scale=0.5]{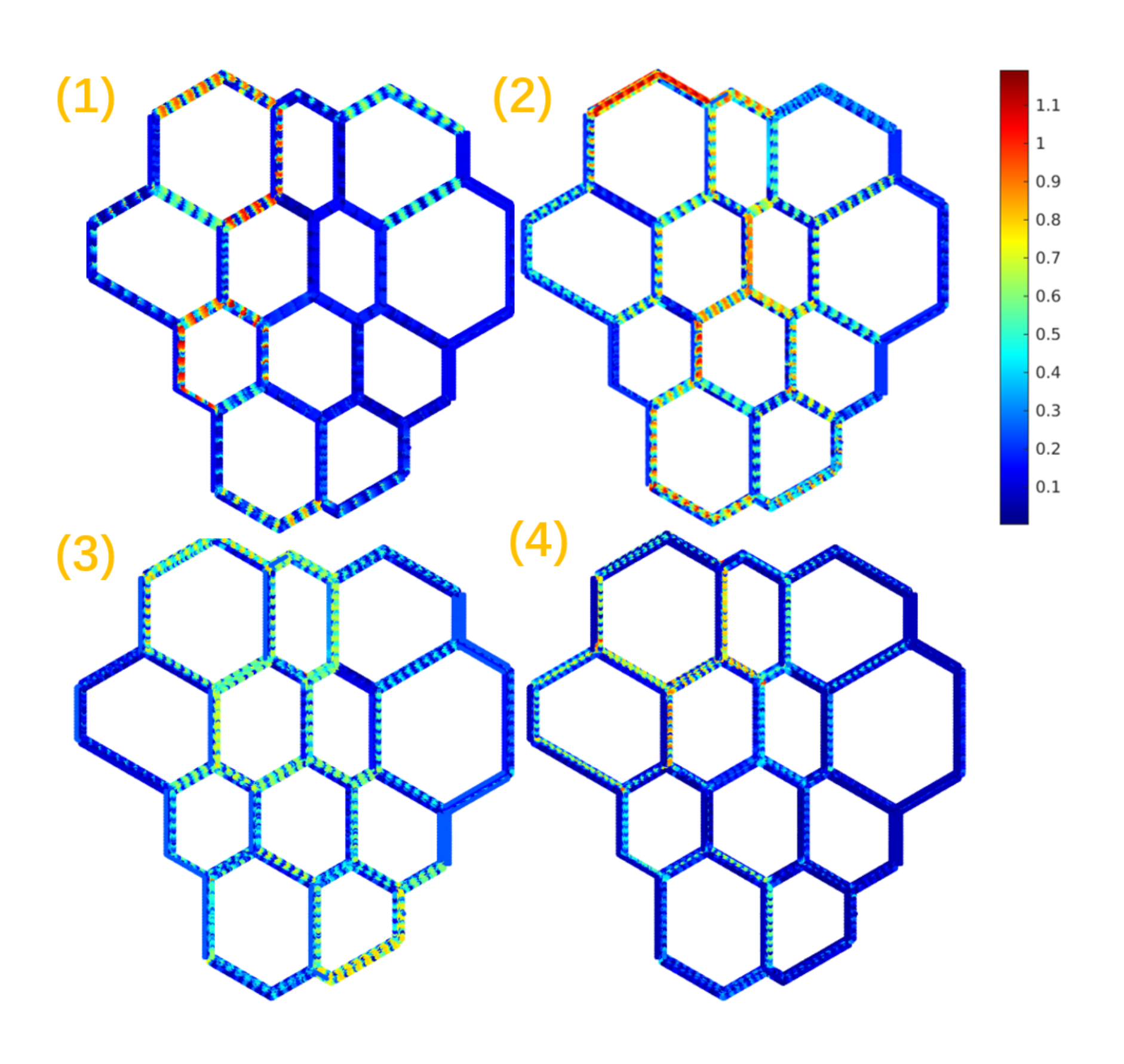}  
\caption{Measured electric field intensities. For the measurement, the microwave frequency was tuned to the eigenfrequencies (1) $8.2202$~GHz, (2) $9.0736$~GHz, (3) $10.0031$~GHz and (4) $11.9859$~GHz. The color scale is indicated in the bar to the right.}   
\label{fig:waf}   
\end{figure}

We measured the electric field intensity distributions, i.e., wave-function intensities, for 154 well isolated resonances in the frequency range $f_{\rm TM_{10}}\leq f\leq f_{\rm TM_{20}}$ where only one mode is excited in transversal direction of the waveguide, using the ports to couple in microwaves. We tuned their frequency to one of the corresponding eigenfrequencies and employed the method explained in~\refsec{MWN}. In order to avoid frequency shifts due to temperature drifts the room temperature was kept constant with an air conditioner. The perturbation body was moved along three different loops, shown in~\reffig{Loops} (1)-(3) along seven straight lines parallel to the waveguide walls, shown schematically in~\reffig{Loops} (4). 
Thus, some of the waveguides were visited more than once. Then we averaged over the intensities resulting from the different loop measurements. Four examples of measured wave functions are shown in~\reffig{fig:waf}. Along the straight waveguide parts, i.e., the bonds, the wave function patterns exhibit sinusoidal oscillations with a constant amplitude. Their transport properties at the junctions are as illustrated in~\reffig{fig:spara_3}. They lead to the complex structure of the intensity which can be vanishingly small in some of the waveguides.  

\begin{figure}[htbp]
\includegraphics[width=0.7\linewidth]{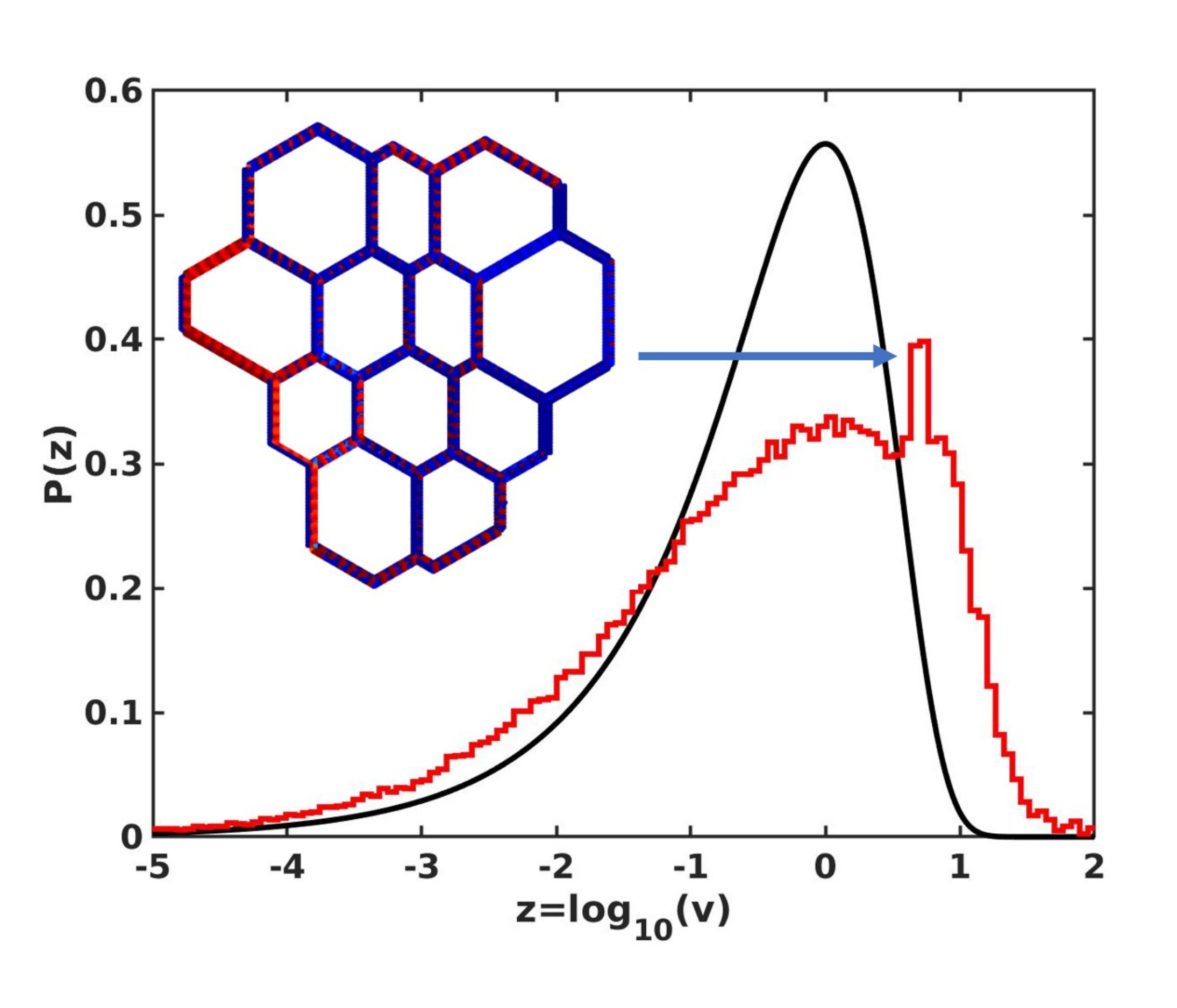}  
\includegraphics[width=0.7\linewidth]{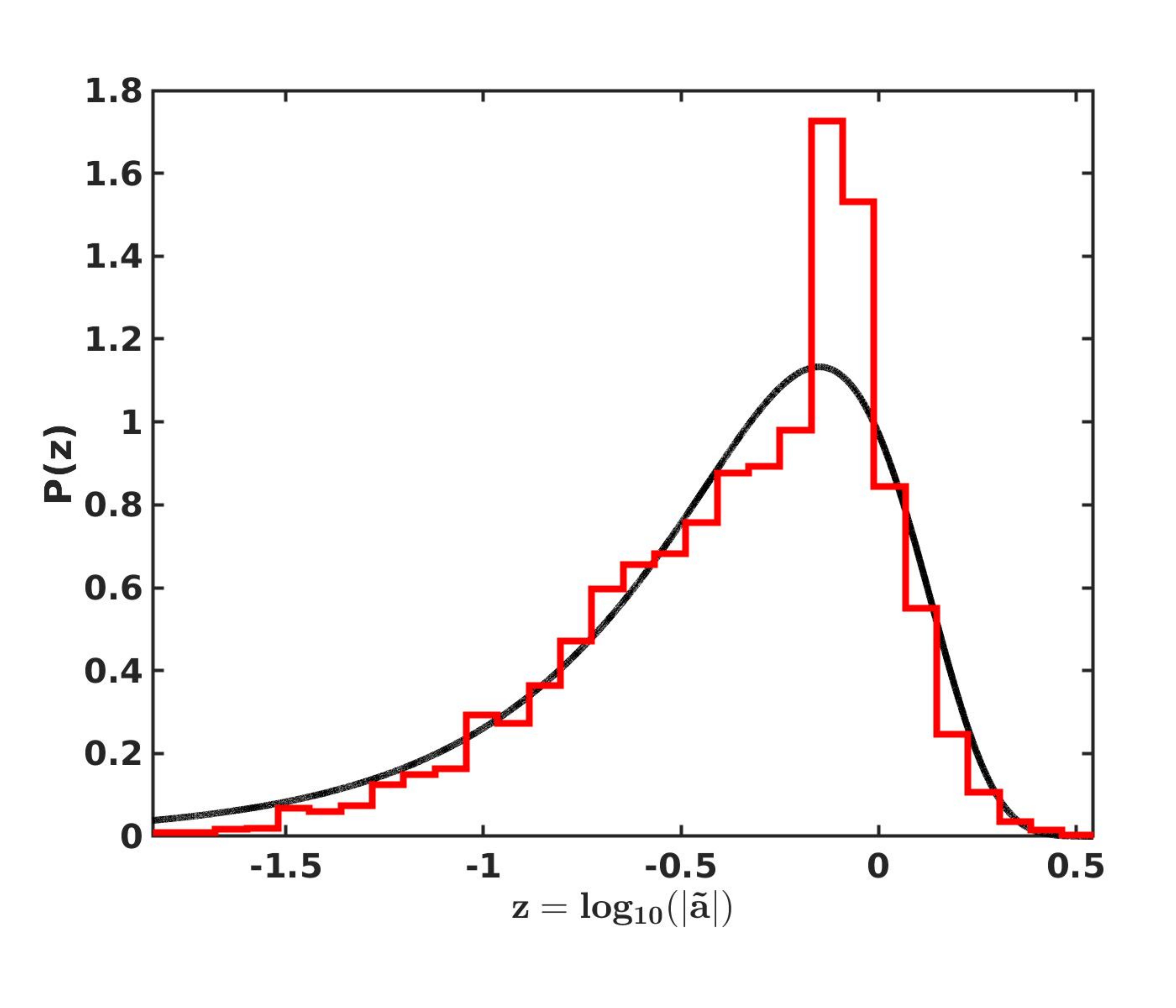}  
\includegraphics[width=0.7\linewidth]{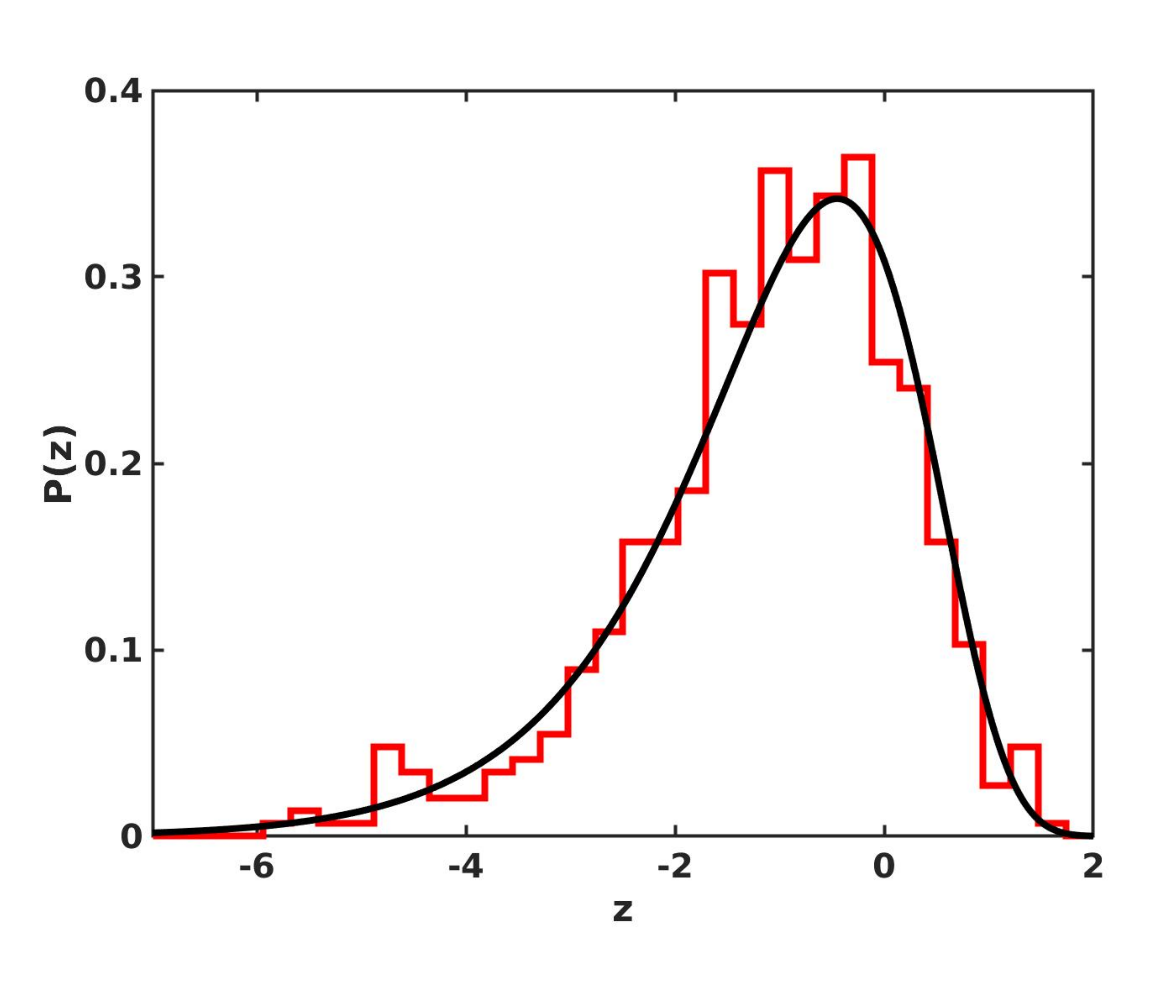}  
\caption{Top: Distribution of the wave-function intensities after normalization as in~\refeq{Norm1}. Middle: Distribution of the wave-function amplitudes in the individual waveguides after normalization as in~\refeq{Norm2}. Bottom: Strength distribution. The red histograms show the results deduced from the experimental data and the black solid line the Porter-Thomas distribution.}    
\label{fig:pwaf}   
\end{figure}

To further explore this structure, we analyzed the distribution of their intensities. For this we employed the experimental data obtained from a measurement along the loop $\mathcal{C}$ of total length $\mathcal{L}_1$ shown in~\reffig{Loops} (2), which comprises in total $N_{\mathcal{B}} =33$ out of the 48 waveguides, in the frequency range of a single transversal mode, $f\leq f_{\rm TM_{20}}$. The top panel in~\reffig{fig:pwaf} exhibits the distribution of the normalized electric field intensity $\vert E_z(l)\vert^2$, with $l$ giving the distance covered by the perturbation body along the loop at each point of measurement,  
\be 
v=\mathcal{N}^{-1}\vert E_z(l)\vert^2,\, \mathcal{N}=\oint_\mathcal{C}\vert E_z(l^\prime)\vert^2\frac{dl^\prime}{\mathcal{L}_1}\label{Norm1}.
\ee
According to the random-plane wave hypothesis~\cite{OConnor1987,Berry2002} the distribution $P(v)$ of the thus normalized squared wave function components is expected to coincide with a Porter-Thomas distribution $P(v)=\frac{1}{\sqrt{2\pi v}}e^{-v/2}$ with mean value unity for quantum systems with a fully chaotic classical dynamics. It has a singularity at $v=0$. Therefore, we transformed $v$ to the logarithmic variable $z=\log_{10}(v)$. The result is exhibited in the top panel of~\reffig{fig:pwaf} (red histogram). Deviations from the Porter-Thomas distribution (black solid line) result from wave functions exhibiting above-average intensities with values beyond $z\simeq 1$. These are localized on a few waveguides. One example is depicted in the inset. 

As illustrated in~\reffig{fig:waf} the wave-function components on the bonds $j=1,\dots,N_{\mathcal{B}}$ are well described for each eigenwavenumber $k_m$ by an ansatz of the form $\psi_j(y)=a_j(x;k_m)e^{ik_my}+a^\ast_j(x;k_m)e^{-ik_my}$ with $-w/2\leq x\leq w/2$, $0\leq y\leq L_j$ and $L_j$ denoting the lengths of the bonds. Accordingly, we proceeded as in~\cite{Kaplan2001} and analyzed wave-function intensities in terms of the distribution of the squared modulus of the amplitudes $a_j(x;k_m)$, which are complex numbers with the star denoting complex conjugation. We determined the amplitudes by identifying the maxima in the electric-field intensity along the central line $x=0$ of the waveguides. Similar to~\refeq{Norm1} we introduce normalized amplitudes $\tilde a_j(k_m)$, 
\be\label{Norm2}
\sum_{j=1}^{N_{\mathcal{B}}} \frac{L_j}{\mathcal{L}_1}|\tilde a_j(k_m)|^2 =1,
\ee
for the computation of the wave-function intensity distribution. Here, we suppress the argument $x=0$. Essentially, in distinction to the procedure~\refeq{Norm1}, this one neglects the $y$-dependence of the electric-field intensity, assuming that it takes its maximal value along the whole bond. For quantum systems with fully chaotic counterpart the squared amplitudes $\vert\tilde a_j\vert^2$ are expected to be Porter-Thomas distributed. In the middle panel of~\reffig{fig:pwaf} we compare the experimental result (red histogram ) with the Porter-Thomas distribution (black solid line). Like in the distribution of $v$ the deviations may be attributed to wave functions that are strongly localized on a few waveguides, thus yielding the exceptionally high peak. 
These results for the distributions of the squared wave-function components $v$ and their amplitudes $\vert\tilde a\vert^2$ suggest that for a nonnegligible part of the waveguide modes the complexity introduced at the junctions of the waveguide network is not sufficient to generate the random-plane wave behavior expected for quantum systems with a fully chaotic classical counterpart. These are localized on part of the waveguide network and thus do not comply with the random-plane wave hypothesis. 

Another possibility to extract information on statistical properties of the wave function components is to exploit the proportionality of the partial widths associated with antennas $a$ and $b$ to the electric field intensities at the positions of the antennas~\cite{Alt1995,StoeckmannBuch2000}. The partial widths cannot be determined separately~\cite{Dembowski2005}. Therefore, we determined the strengths $y=\gamma_{\mu a}\gamma_{\mu b}$ by fitting~\refeq{Eq:bw} to the transmission spectra and analyzed their distributions. We rescaled them to average value unity, $\langle\tilde y\rangle=1$. In the bottom panel of~\reffig{fig:pwaf} we compare the distribution of the transformed strength, $z=\log_{10}\widetilde{y}$, with that expected for quantum systems with a fully chaotic classical counterpart, which is given as~\cite{Dembowski2005}
\begin{equation}
p(z)=\frac{\ln(10)}{\pi}10^{z/2}K_0(10^{z/2}).
\end{equation}
Here, $K_0(x)$ denotes the modified Bessel function of order zero~\cite{Abramowitz2013}. The experimental results were obtained by averaging over the distributions for three antenna combinations. Agreement with the RMT prediction is good. In distinction to the wave function measurements, only the wave-function components at 8 positions are taken into account at all eigenfrequencies. There, the wave-function intensities are not necessarily maximal, however they are non-zero, because otherwise the resonance would be missing in the transmission spectra.

\begin{figure*}[!ht]
\includegraphics[scale=0.42]{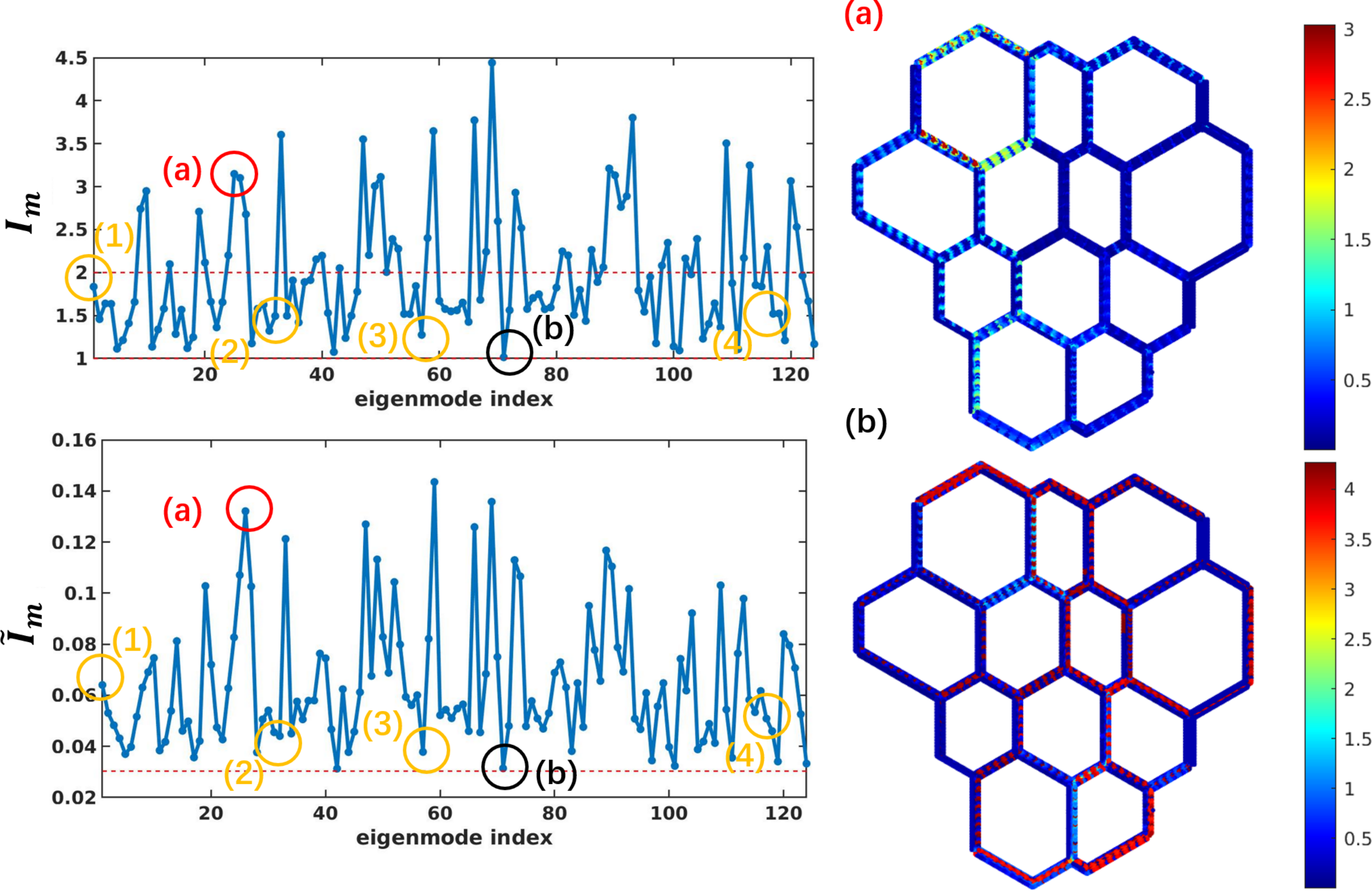}
\caption{Inverse participation ratio of the eigenmodes of the waveguide network computed according to~\refeq{IPR1} (top) and~\refeq{IPR2} (bottom). The right part shows two further examples for measured electric-field intensities.  The color scales are indicated in the bars to the right. The IPR values corresponding to these examples and to those exhibited~\reffig{fig:waf} are marked by the red, black and yellow circles, respectively.}   
\label{fig:ipr}    
\end{figure*}
The inverse participation ratio (IPR), 
\be\label{IPR1}
\mathcal{I}_m=\langle\vert\tilde a_j(k_m)\vert^4\rangle
\ee
provides an additional statistical measure for the wave-function intensity distribution applied in Ref.~\cite{Kaplan2001} to quantum graphs, assuming that the amplitudes $a_j(k_m)$ are normalized as in~\refeq{Norm2}, which corresponds to the first non-trivial moment of the distribution $P(\vert\tilde a\vert^2)$ introduced above. The IPR gives information on the degree of localization of a wave function~\cite{Kaplan2001,Hul2009}, i.e., the degree of deviation of it from a random plane wave. In the limit of maximal ergodicity in configuration space, where all intensities $\vert a_j\vert$ are equal, $\mathcal{I}=1$. The other extremal situation corresponds to localization on individual bonds (b), where it equals their number, $\mathcal{I}=N_{\mathcal{B}}$. This situation occurs in quantum graphs with Dirichlet boundary conditions at the vertices~\cite{Kottos1999}. Since the coefficients entering the random plane-wave ansatz are complex, RMT predicts $\mathcal{I}=2$ for time-reversal invariant quantum systems with a fully chaotic classical dynamics. As shown in the top panel in~\reffig{fig:ipr}, the IPR values vary erratically with the eigenfrequency. Values of $N_\mathcal{B}>\mathcal{I}\geq 2$ indicate localization in parts of the waveguide network. In the right part of the figure are shown two examples of measured electric-field intensities, one which exhibits a complex intensity patter on a fraction of the whole waveguide graph (a), and one for which the intensities are similar on all bonds. The corresponding values are larger than two and close to one respectively, as expected. The orange circles indicate the IPR values for the wave functions shown in~\reffig{fig:waf}. The wave function marked by (1) has a value close to the RMT prediction $\mathcal{I}\simeq 2$ and the others a value $\mathcal{I}\simeq 1.5$ between the ergodic and RMT cases. The average over all IPR values equals $\langle\mathcal{I}\rangle=1.9724$ which is close to the RMT value. In~\cite{Hul2009} the IPR is defined as
\be\label{IPR2}
\mathcal{\tilde I}_m=\frac{\sum_{j=1}^{N_\mathcal{B}}\vert a_j(k_m)\vert^4}{\left[\sum_{j=1}^{N_\mathcal{B}}\vert a_j(k_m)\vert^2\right]^2},
\ee
which corresponds to a normalization of $\vert a_j(k_m)\vert^2$ to mean value unity. It equals $\mathcal{\tilde I}=\frac{1}{N_\mathcal{B}}$ for the ergodic case, $\mathcal{\tilde I}=\frac{2}{N_\mathcal{B}}$ for the random-plane wave case and $\mathcal{\tilde I}=\frac{1}{N_L}$, if the wave function is localized on $N_L$ individual bonds. The resulting IPR values are shown in the bottom panel. They reflect the features of the wave functions as demonstrated for the wave function shown in Figs.~\ref{fig:waf} and~\ref{fig:ipr}. The average value $\langle\mathcal{\tilde I}\rangle=0.0645$ is closer to that for the RMT case, $\mathcal{\tilde I}=2/33=0.0606$, than to that for the ergodic case and well below that for localized wave functions.

\subsection{Fluctuation properties in the eigenfrequency spectrum above the cutoff frequency for two transversal modes}
\begin{figure}[htbp]
\includegraphics[width=0.9\linewidth]{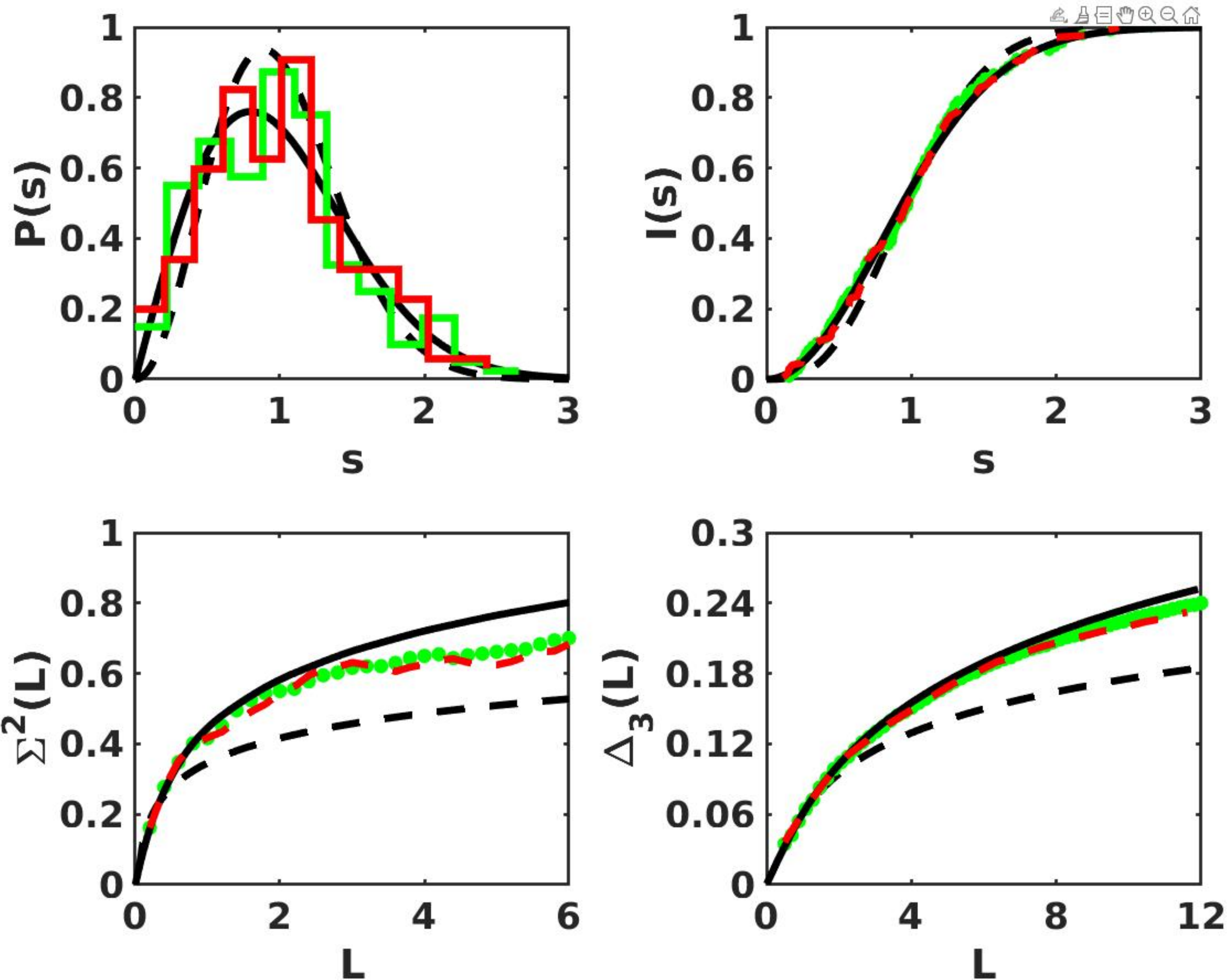}
\caption{Same as~\reffig{fig:sps1d} for the frequency range above the cutoff frequency of the second transversal TM$_{20}$ modes. Shown are the results deduced from the experimental data (red histogram and dashed lines) and the COMSOL Multiphysics simulations (green line and dots)}   
\label{fig:sps2d}  
\end{figure}

We measured transmission and reflection spectra up to the cutoff frequency $f_{\rm TM_{01}}$ for the first excited transverse magnetic mode which also comprises eigenfrequencies $f_m\geq f_{\rm TM_{20}}$ corresponding to the second transversal mode. For these eigenmodes wave propagation is no longer one-dimensional and the analogy to the quantum graph of corresponding geometry is lost. We identified a complete sequence of 176 eigenfrequencies in the range from 13.5 to 14.7 GHz. Furthermore, we computed the eigenfrequencies in that range employing COMSOL Multiphysics. The spectral statistics deduced from the experimental and numerical data are shown in~\reffig{fig:sps2d}. 
Their agreement with those of random matrices from the GOE is similar to that for the frequency range of one transversal mode shown in~\reffig{fig:sps1d}. 

\section{Fluctuation properties in the eigenfrequency spectrum with a single transversal mode in the presence of partial \Ti-invariance violation\label{TIV}}
\begin{figure}[htbp]
\includegraphics[width=0.9\linewidth]{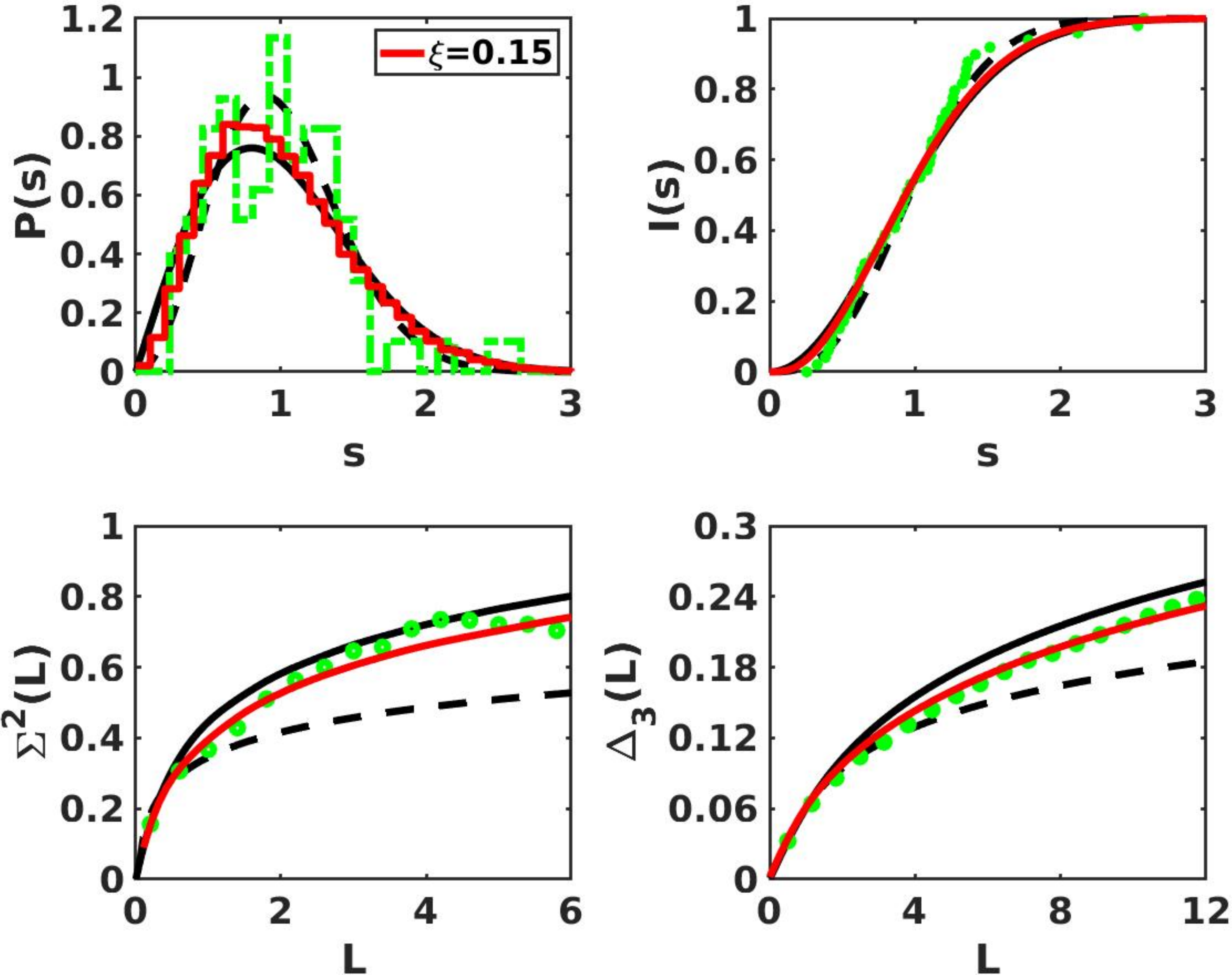}  
\includegraphics[width=0.9\linewidth]{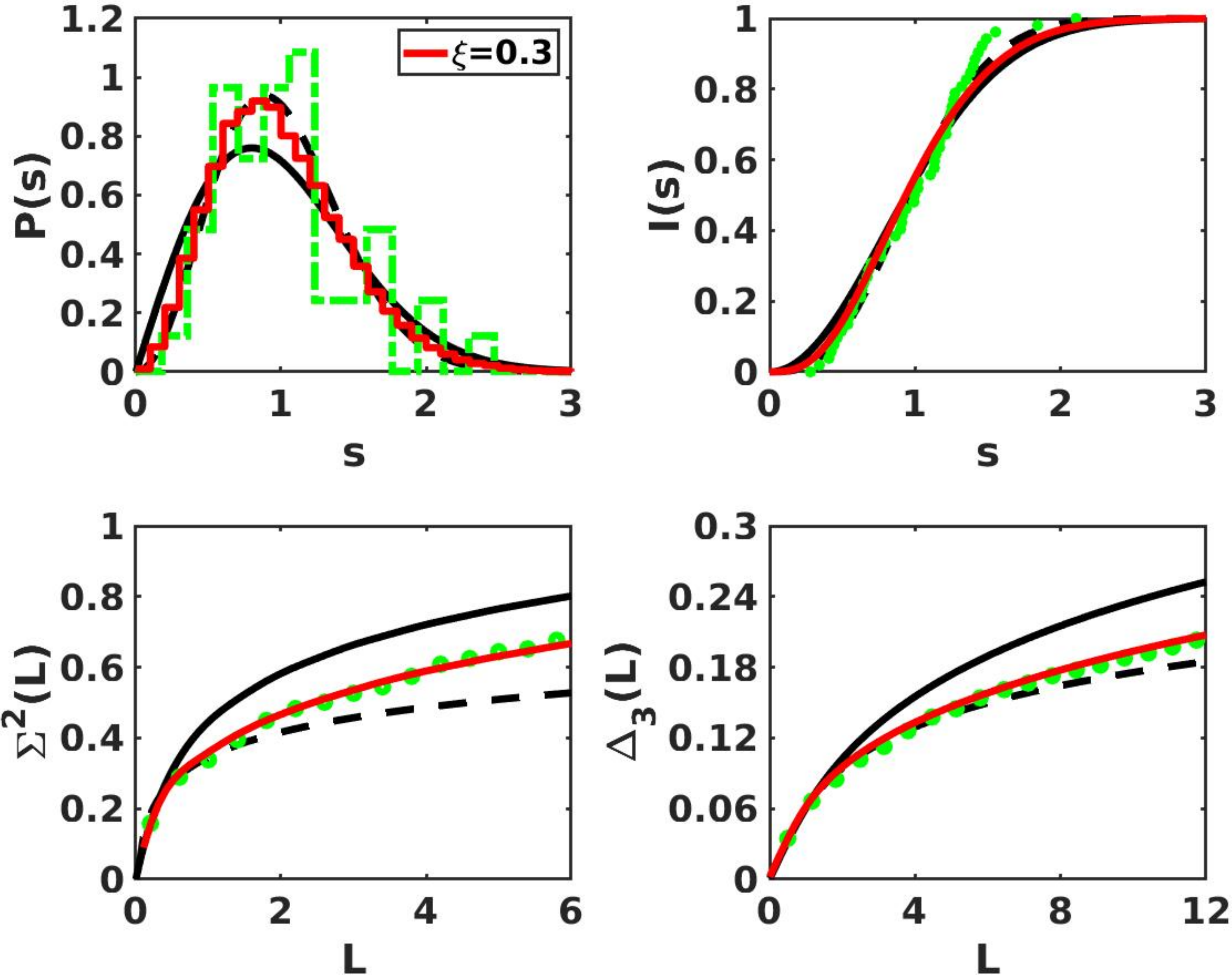}  
	\caption{Spectral properties of the waveguide graph in the frequency ranges 7.1-9.6~GHz (top) and 8.7-14.5~GHz (bottom), respectively, deduced from the experimental data (green dashed lines and dots). The red lines show the curves obtained from the corresponding analytical results Eqs.~(\ref{ps}) and~(\ref{yl}). The strength of \Ti-invariance violation $\xi$ is indicated in the inset.}    
\label{fig:spsfe}   
\end{figure}

Quantum systems with partially violated time-reversal invariance and a classically chaotic counterpart are described within RMT by random matrices interpolating between GOE and GUE~\cite{Pandey1981,Altland1993}, 
\begin{equation}
        H_{\rm\mu\nu}=H_{\rm\mu\nu}^{(S)}+i\frac{\pi\xi}{\sqrt{N}}
        H_{\rm\mu\nu}^{(A)}.
\label{eqn:hamiltonian}
\end{equation}
Here, $\hat H^{(S)}$ is a real-symmetric random matrix drawn from the GOE and $\hat H^{(A)}$ is a real-antisymmetric matrix with Gaussian-distributed matrix elements. The parameter $\xi$ determines the magnitude of \T violation. For $\xi =0$ $\hat H$ describes chaotic systems with preserved \T invariance, whereas for $\pi \xi / \sqrt{N} = 1$ $\hat H$ is a random matrix from the GUE, however, the transition from GOE to GUE already takes place for $\xi\simeq 1$~\cite{Dietz2010}. Analytical expressions exist for the nearest-neighbor spacing distribution $P(s;\xi)$ and the two-point cluster function $Y_2(L;\xi)$. They are given by~\cite{Lenz1992}
\be
P(s;\lambda)=\sqrt{\frac{2+\lambda^2}{2}}sc^2(\lambda){\rm erf}\left(\frac{sc(\lambda)}{\lambda}\right)e^{-\frac{s^2c(\lambda)^2}{2}}
\label{ps}
\ee
with $\lambda=\frac{\pi}{\sqrt{2}}\xi$, 
\be c(\lambda)=\sqrt{\pi\frac{2+\lambda^2}{4}}\left[1-\frac{2}{\pi}\left(\tan^{-1}\left(\frac{\lambda}{\sqrt{2}}\right)-\frac{\sqrt{2}\lambda}{2+\lambda^2}\right)\right]
\ee
and ${\rm erf}(x)$ denoting the error function, and~\cite{Mehta1990,Pandey1991,Bohigas1995}
\begin{equation}
Y_2(L;\xi)=\det\left[
\begin{pmatrix}
s(L) &-D(L;\xi)\\
-J(L;\xi) &s(L)
\end{pmatrix}
\right],
\label{yl}
\end{equation}
with $s(L)=\frac{\sin\pi L}{\pi L}$, $D(L;\xi)=\frac{1}{\pi}\int_0^\pi{\rm d}xe^{2\xi^2 x^2}x\sin(Lx)$ and $J(L;\xi)=\frac{1}{\pi}\int_\pi^\infty{\rm d}xe^{-2\xi^2x^2}\frac{\sin(Lx)}{x}$. The number variance $\Sigma^2(L;\xi)$ and spectral rigidity $\Delta_3(L;\xi)$ are obtained from the two-point cluster function~\cite{Mehta1990,Bialous2021}. The red curves in~\reffig{fig:spsfe} are deduced from these analytical expressions for the values of the parameter $\xi$ indicated in the panels. To determine it we proceeded as in~\cite{Dietz2009a,Dietz2010} using the scattering matrix formalism for microwave resonators as outlined in the next section, ~\refsec{Smatrix}.

For the measurement of the transmission and reflection spectra of the waveguide graph containing the ferrites, that are magnetized with external magnets as described in~\refsec{MWN}, only antennas at the positions marked by 1,2,7 and 8 were used which are located outside the region of the ferrites. Strongest \Ti-invariance violation is observed for the single-mode case in the frequency range 10-12~GHz, and above the cutoff frequency for the second transversal mode in the range 13-14~GHz. The spectral properties are presented in~\reffig{fig:spsfe}, where we combined the results for these frequency ranges. We show results for the ranges from 7.1-9.6~GHz and  8.7-14.5~GHz. Panels (b) and (d) of~\reffig{fig:pr} show the ratio distribution and cumulative ratio distribution of the non-unfolded eigenfrequencies in the frequency range 8.7-14.5~GHz. For the nearest-neighbor spacing distribution and its cumulative distribution differences between this model and the GOE and GUE curves are clearly visible for $\xi =0.15$, whereas they are close to those of the GUE for  $\xi =0.3$. The ratio distribution is close to that for the GUE for both values of $\xi$. They are shown for  $\xi =0.3$ in~\reffig{fig:pr}.
The spectral properties agree quite well with those of the random-matrix model~\refeq{eqn:hamiltonian}.

\section{Fluctuation properties of the scattering matrix\label{Smatrix}}
We also analyzed to what extent waveguide graphs exhibit features typical for quantum-chaotic scattering systems. Since the reflection and transmission spectra are measured by emitting microwave power into the resonator via one antenna or port -- where, according to the results obtained for the wave functions (see, e.g.~\reffig{fig:waf}) the microwaves experience mainly in the regions of the junctions scattering from the resonator walls -- and then receiving them at the same or another antenna or port, the waveguide graph can be viewed as a scattering system~\cite{Albeverio1996}. The associated scattering matrix is given in~\refeq{eq:SResonator}. The scattering formalism is identical with that introduced in~\cite{Mahaux1969} for the description of compound-nucleus reactions. This analogy has been employed in numerous experiments~\cite{Bluemel1990,Mendez-Sanchez2003,Schaefer2003,Kuhl2005,Hul2005,Dietz2008,Dietz2009a,Dietz2010,Dietz2010a,Dietz2011a} to investigate universal properties of the scattering matrix for compound-nucleus reactions and, generally, for quantum scattering processes with classically chaotic dynamics and preserved or partially violated \T invariance. Analytical results~\cite{Verbaarschot1985,Pluhar1995} were derived on the basis of the supersymmetry and RMT approach and verified in Refs.~\cite{Dietz2008,Dietz2009a,Dietz2010}  for scattering-matrix correlation functions and in Refs.~\cite{Fyodorov2005,Kumar2013,Kumar2017} for distributions of the scattering-matrix elements.

Within this RMT approach for quantum-chaotic scattering systems with partially violated \T invariance, the scattering matrix is obtained by replacing in~\refeq{eq:SResonator} the resonator Hamiltonian by a random matrix of the form~\refeq{eqn:hamiltonian}. Furthermore, the matrix $\hat W$ accounts for the coupling of the $N$ internal modes to their environment through $M$ open channels modeling the antennas or ports and $\Lambda$ fictitious channels~\cite{Verbaarschot1986,Savin2006} that mimick the absorption into the walls of the resonator. Thus it is a $(M+\Lambda)\times N$ dimensional matrix with real and Gaussian distributed entries $W_{c\mu}$ of which the sum $\sum_{\mu=1}^NW_{c\mu}W_{c\mu}= Nv_c^2,\, c=1,\dots,M+\Lambda$ yields the transmission coefficients 
\begin{equation}\label{eq:transcoe}
T_c=1-\vert\langle S_{cc}\rangle\vert^2,
\end{equation}
 through the relation $T_c=\frac{4\pi^2v^2_c/d}{(1+\pi^2v^2_c/d)^2}$, with $d=\sqrt{\frac{2}{N}\langle H^2_{\mu\mu}\rangle}\frac{\pi}{N}$ denoting the mean resonance spacing~\cite{Dietz2010}. They provide a measure for the unitarity deficit of the average scattering matrix $\langle S_{cc}\rangle$. The frequency-averaged scattering matrix obtained from the transmission and reflection measurements is diagonal, $\langle S_{cc^\prime}\rangle=0$, implying that direct processes are negligible. This property is accounted for in the RMT model through the orthogonality property $\sum^N_{\mu=1}W_{c\mu}W_{c^\prime\mu} = Nv^2_c\delta_{cc^\prime}$. For indices $c$ denoting an antenna or port channel the parameter $v^2_c$ corresponds to the average size of the electric field at the position of the antenna or port~\cite{StoeckmannBuch2000}. For the RMT simulations we chose an ensemble of 200 random matrices with $M=2,\,\Lambda=30,\, N=300$. 
 
The input parameters of the RMT model for the scattering matrix are the transmission coefficients $T=T_{a}\simeq T_{b}$ associated with antennas or ports $a$ and $b$, the transmission coefficients $T_f\simeq T,\, f=3,\dots,M+\Lambda$ accounting for the absorption $\tau_{abs}=\Lambda T_{f}$ and the \T-violation parameter $\xi$. The transmission coefficients $T_a$ and $T_b$ associated with antennas $a$ and $b$ are obtained with \refeq{eq:transcoe} from the measured reflection spectra, whereas preliminary values for the absorption parameter $\tau_{abs}$ are obtained by fitting the complex Breit-Wigner form \refeq{Eq:bw} to the measured resonances and employing the Weisskopf formula~\cite{Blatt1952},
\begin{equation}\label{eq:Weisskopf}
2\pi\frac{\Gamma}{d}=\sum_cT_c=T_1+T_2+\tau_{abs}.
\end{equation}
This value for $\tau_{abs}$ is further refined by proceeding as in~\cite{Dietz2010} and comparing distributions of the experimental scattering-matrix elements $S_{ab},\, a,b=1,2$, and the autocorrelation function
\begin{equation}\label{eq:autocor}
C_{ab}(\varepsilon)=\langle S^{\rm fl}_{ab}(f)S^{\ast\rm fl}_{ab}(f+\varepsilon)\rangle
\end{equation}
with $S^{\rm fl}_{ab}(f)=S_{ab}(f)-\langle S_{ab}(f)\rangle$ and $\langle\cdot\rangle$ denoting the spectral average over a measured resonance spectrum and an ensemble average over the different antenna or port measurements, to the analytical results~\cite{Fyodorov2005,Dietz2009a,Dietz2010,Kumar2013,Kumar2017}. The size of $\xi$ is determined from the cross-correlation coefficients $C^{cross}_{12}(0)=C^{cross}_{12}(\varepsilon=0;\tau_{abs},\xi)$,
\begin{equation}
\label{Eq.3}
C^{cross}_{12}(0)=\frac{\Re[{\langle S^{\rm fl}_{12}(\nu)\, S^{\ast\rm fl}_{21}(\nu)\rangle]}}{\sqrt{\langle|(S^{\rm fl}_{12}(\nu)|^2\rangle\langle|(S^{\rm fl}_{21}(\nu)|^2\rangle}},
\end{equation}
and the corresponding analytical result~\cite{Dietz2009a}. It equals unity for \Ti-invariant systems, and approaches zero with increasing size of \Ti-invariance violation. 

\begin{figure}[htbp]
        \includegraphics[width=0.49\linewidth]{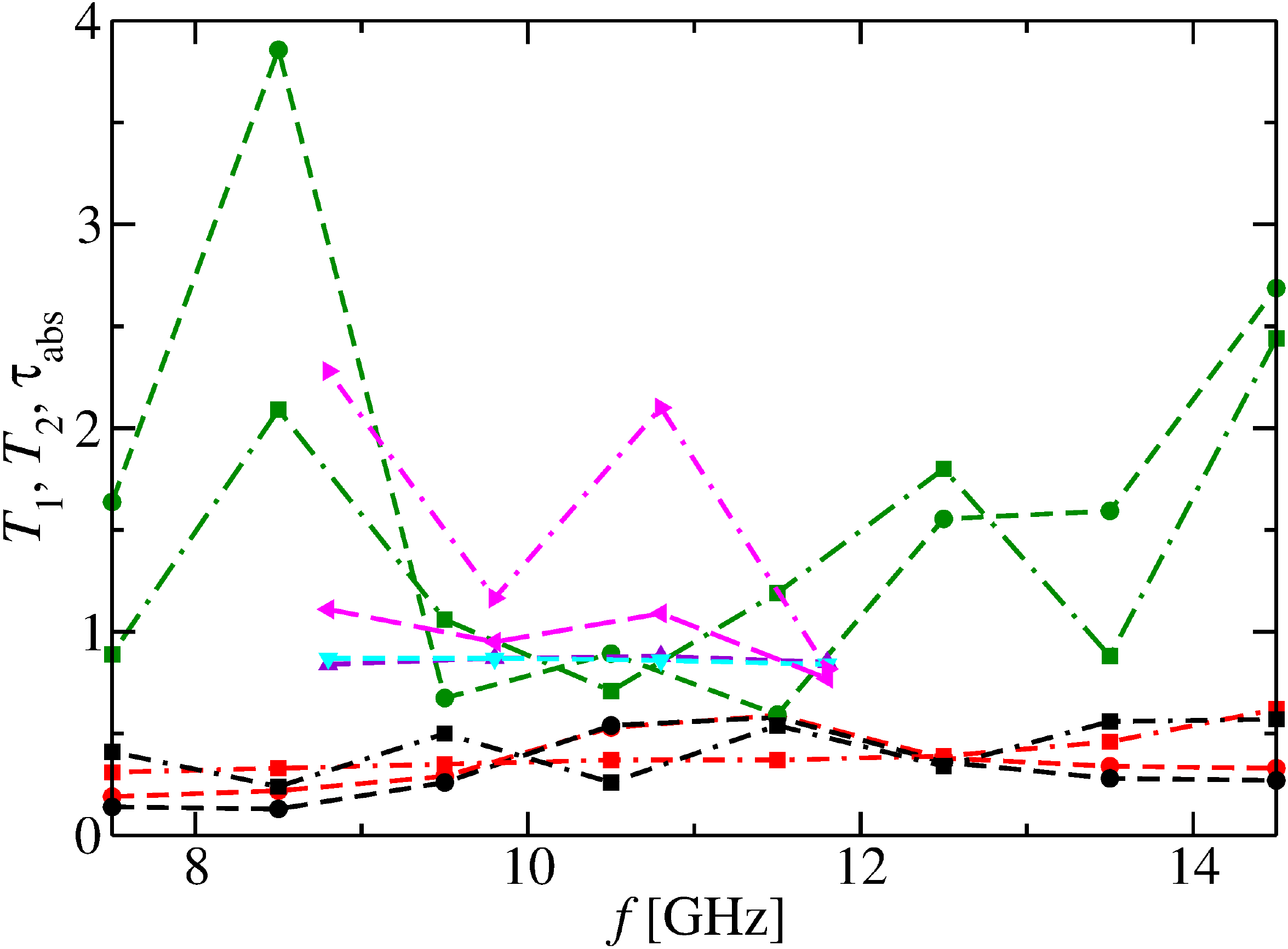}  
        \includegraphics[width=0.49\linewidth]{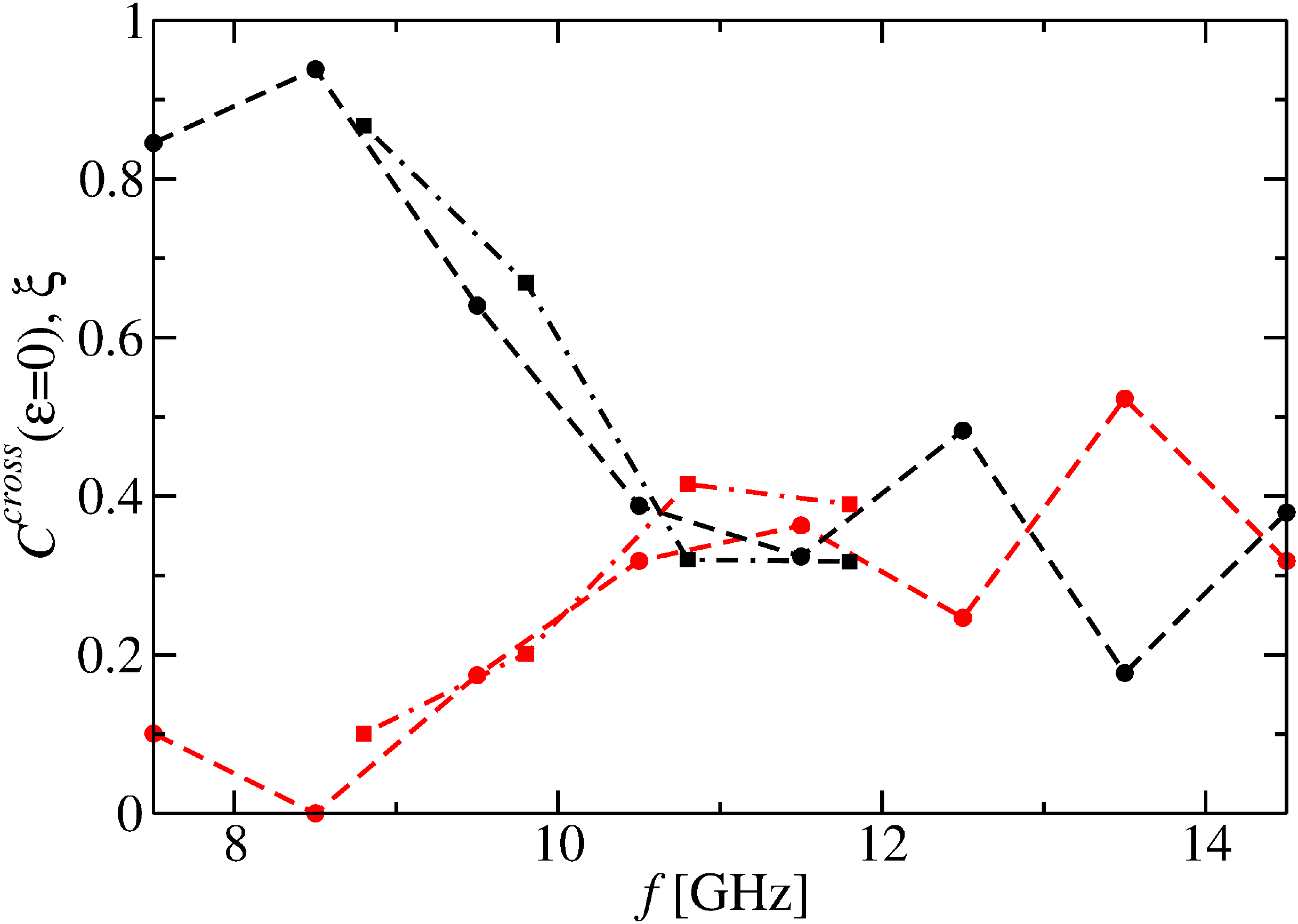}
	\caption{Left: Variation of transmission coefficients $T_1$ (black [cyan]), $T_2$ (red [violet]) and $\tau_{abs}$ (green [magenta]) deduced from the antenna [port] measurements without (dots [triangles down and left]) and with magnetized ferrites (squares [triangles up and right]), obtained from averaging over 1~GHz windows. Right: Cross-correlation coefficients (red) and corresponding values of $\xi$ (black) deduced from the antenna (dots connected by dashed lines) and port (squares connected by dashed-dotted lines) measurements.}    
\label{fig:trancoe}
\end{figure}

\begin{figure}[htbp]
\includegraphics[width=0.49\linewidth]{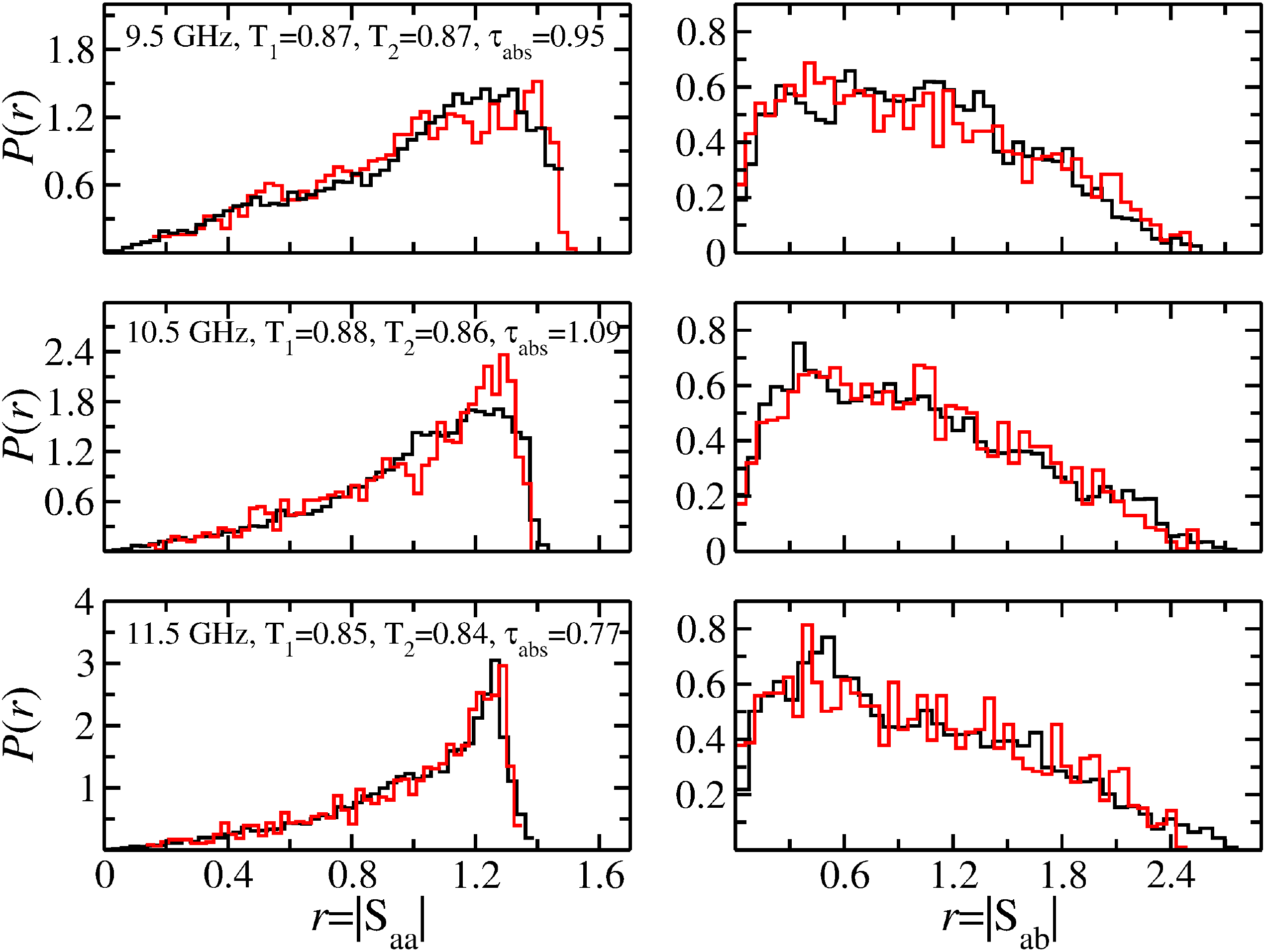}
\includegraphics[width=0.49\linewidth]{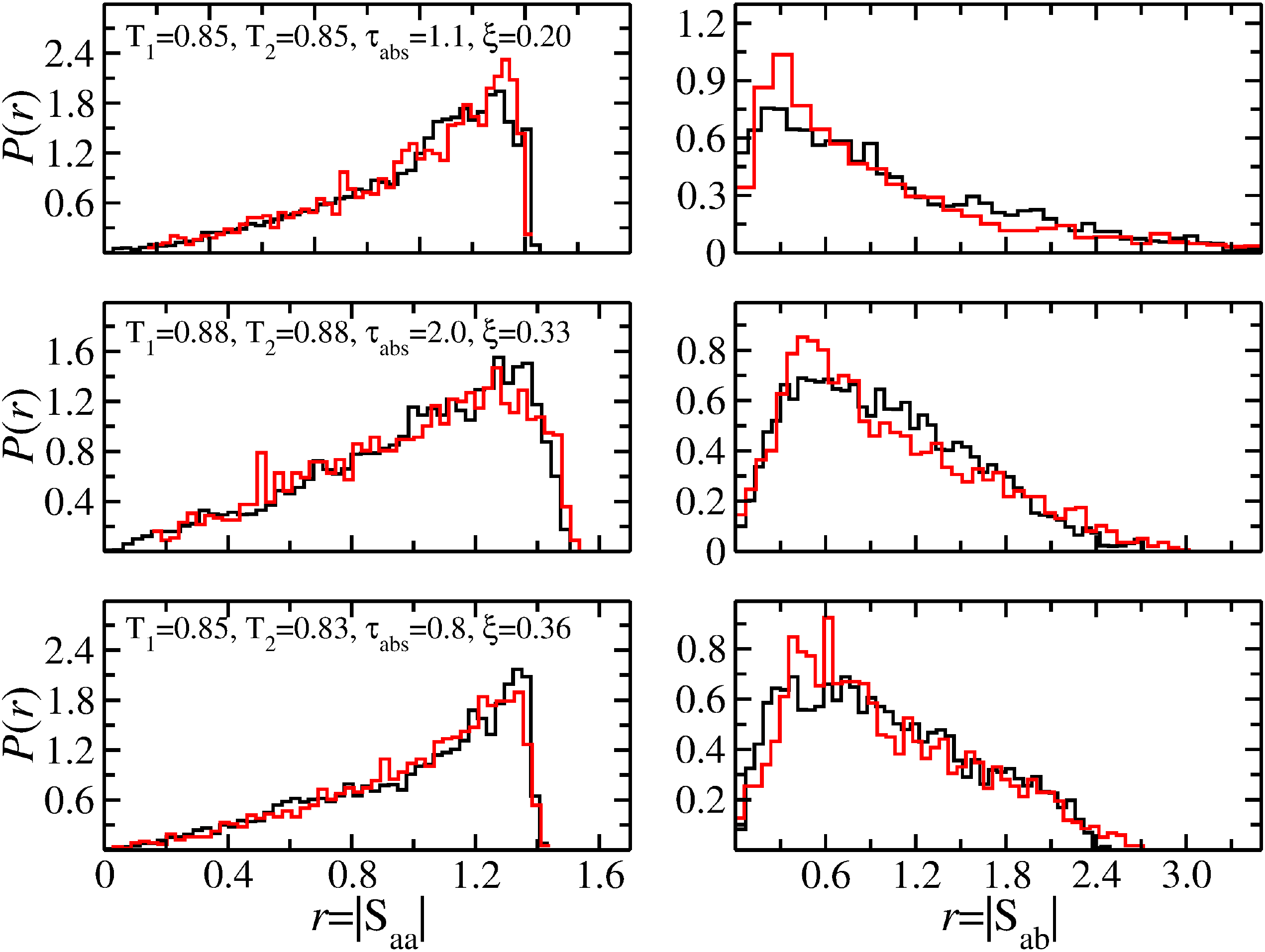}
	\caption{ Distribution of scattering-matrix elements obtained from the port measurements without (left) and with (right) magnetized ferrites (black). They are compared to the corresponding RMT results (red histograms). The corresponding values of $T_1,T_2,\tau_{abs}$ and $\xi$ are given in the panels.}
\label{fig:s12disport}
\end{figure}

\begin{figure}[htbp]
\includegraphics[width=0.241\linewidth]{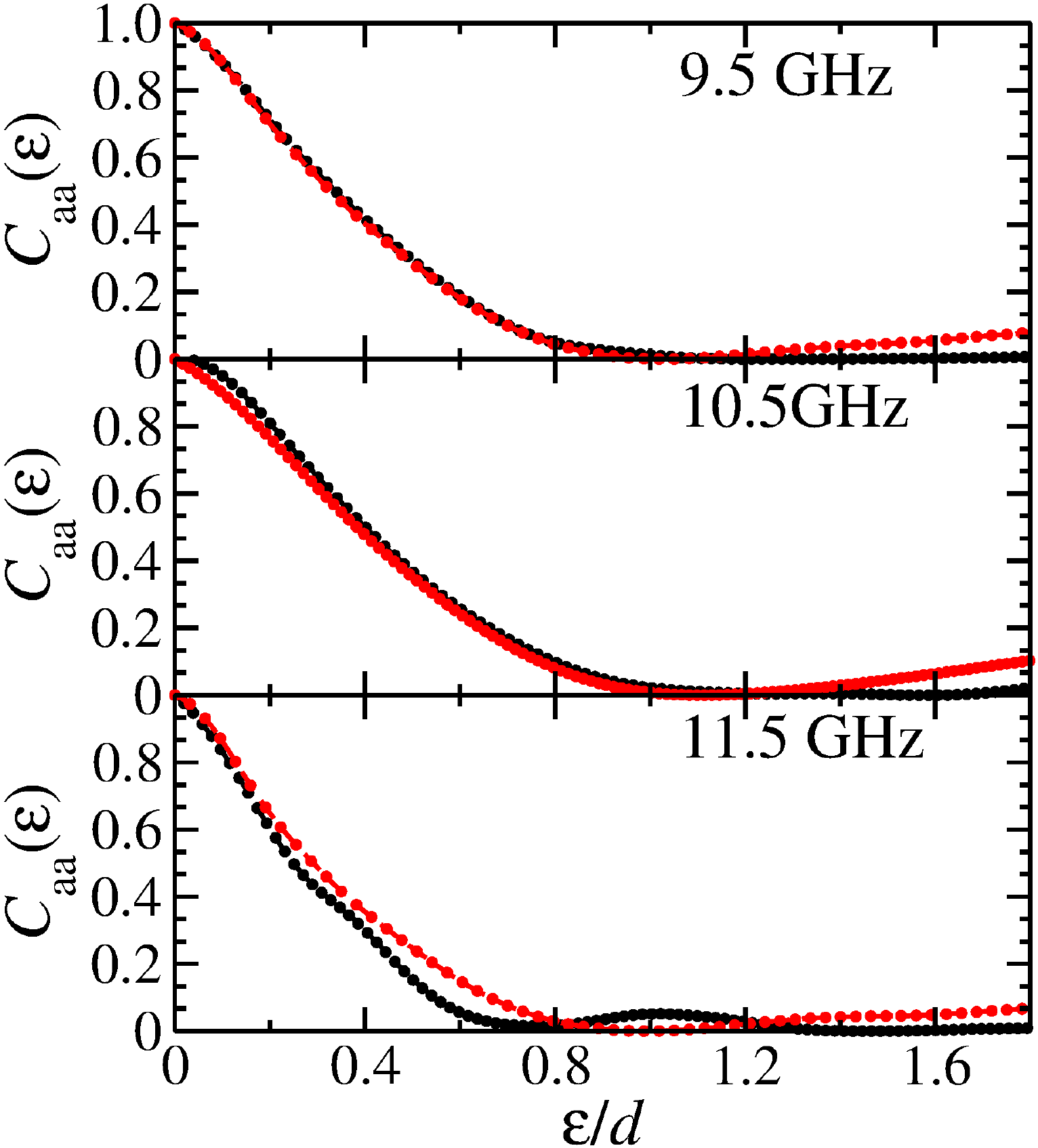}  
\includegraphics[width=0.241\linewidth]{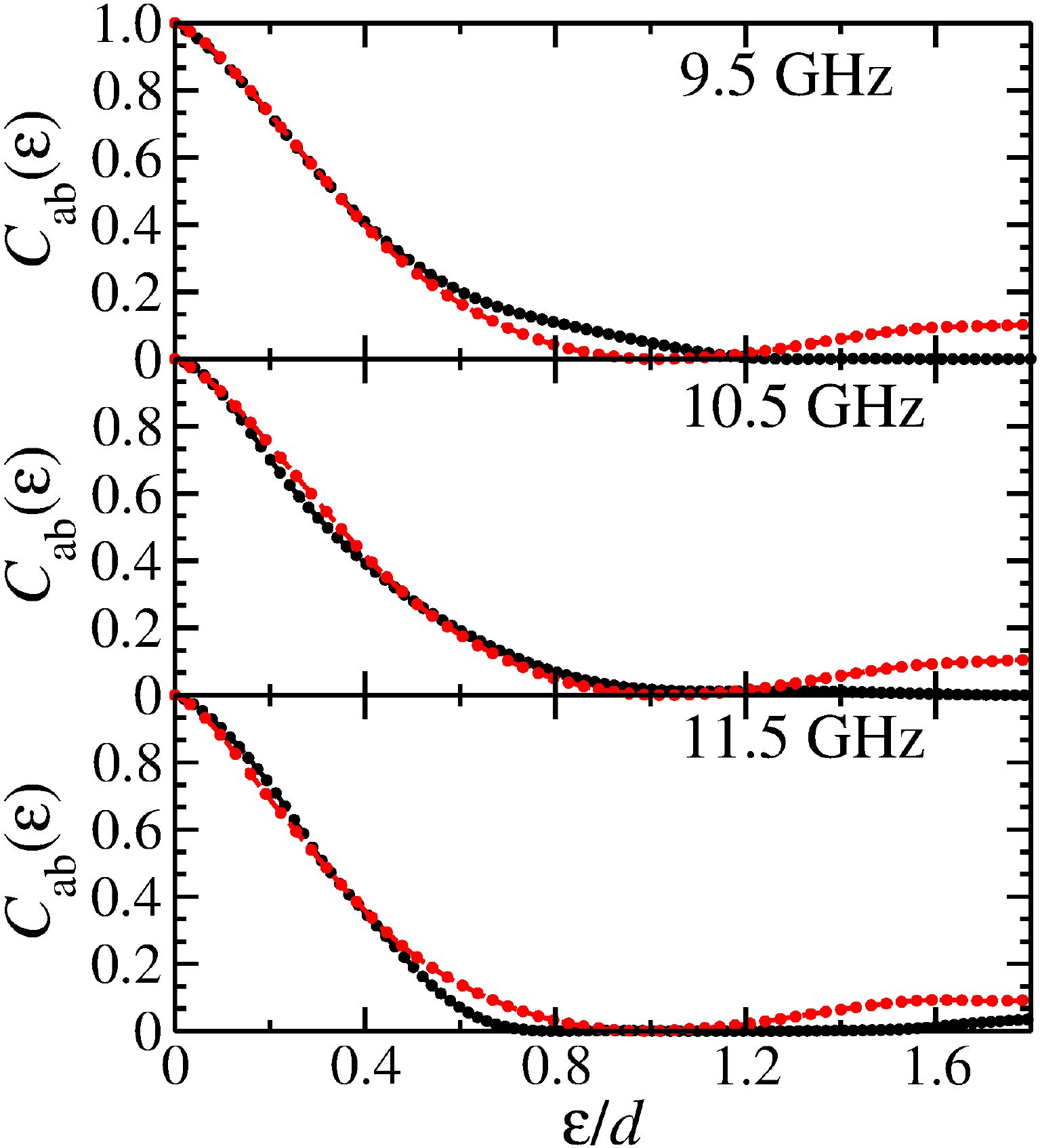}  
\includegraphics[width=0.241\linewidth]{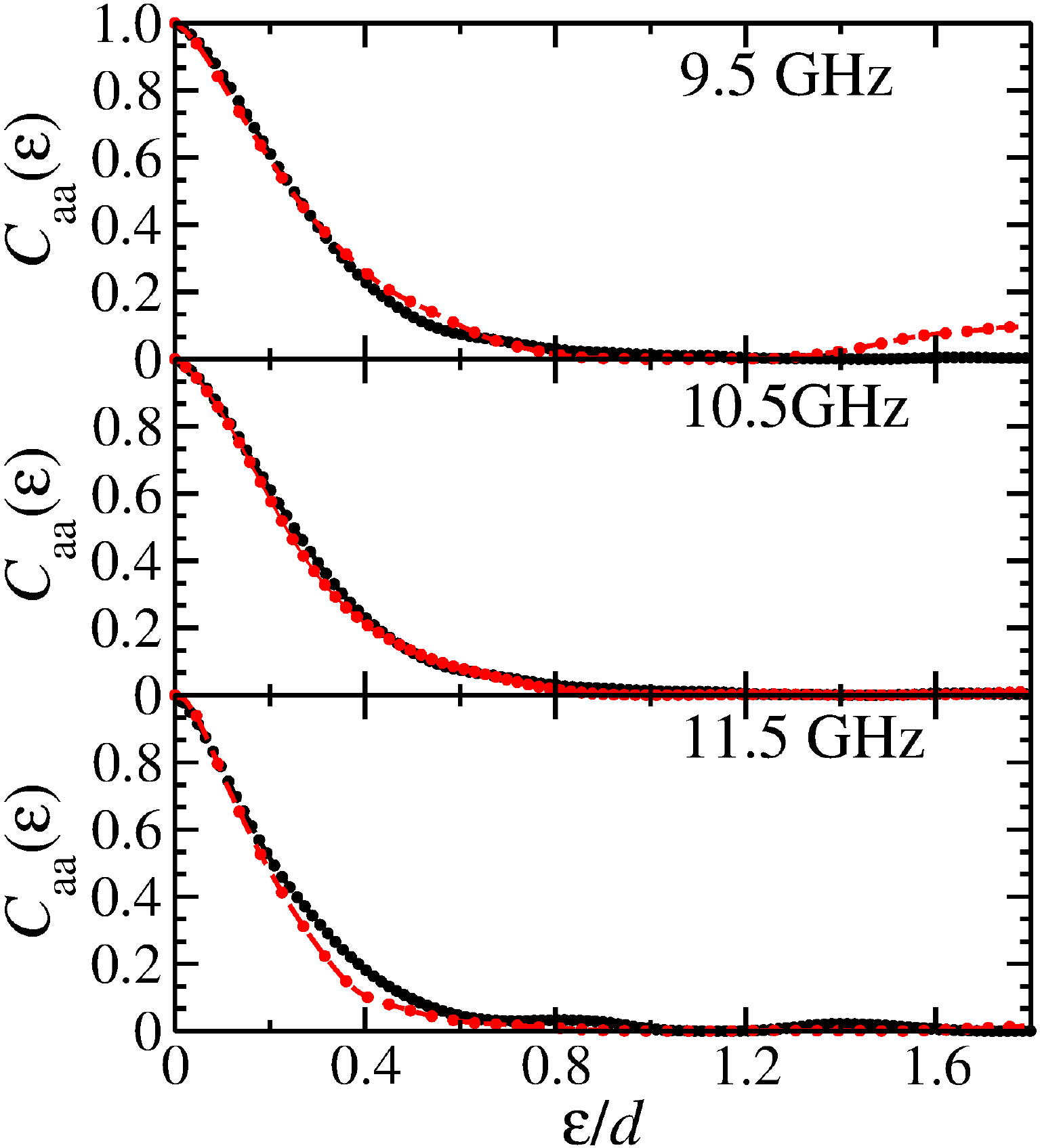}  
\includegraphics[width=0.241\linewidth]{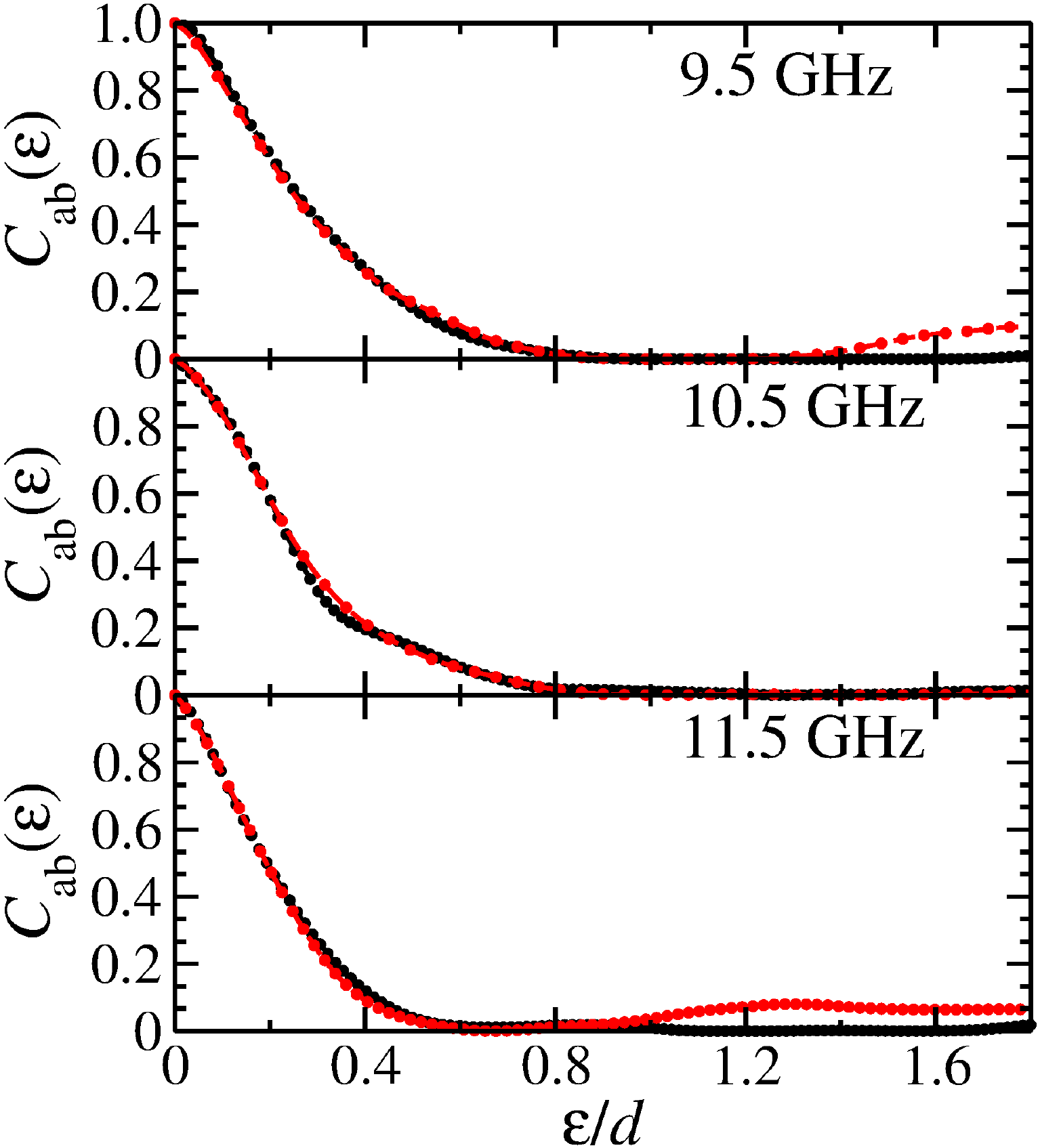}  
\caption{Scattering-matrix autocorrelation functions obtained from the port measurements in the frequency intervals 9-10~GHz, 10-11~GHz and 11-12~GHz without (left) and with magnetized ferrites (right) (black dots and lines). They are compared to the RMT results (red histograms) for the values of $T_1,T_2,\tau_{abs}$ and $\xi$ given in the panels of~\reffig{fig:s12disport}. The frequency difference $\epsilon$ is plotted in units of the local mean level spacing $d$. All curves are normalized to unity at $\epsilon = 0$. }  
\label{fig:corrport}   
\end{figure}

In the RMT model obtained from~\refeq{eq:SResonator} the coupling matrix $\hat{W}$ is assumed to be frequency independent. Therefore, in order to attain approximately frequency-independent resonance parameters, we divided the frequency range into windows of 1~GHz in the analysis of the experimental data~\cite{Dietz2010,Bialous2020}. The results for the transmission coefficients, $\tau_{abs}$ and $\xi$ are shown in~\reffig{fig:trancoe}. For the port measurements the transmission coefficients are of the same size as in the experiments with microwave networks and they are considerably larger than for the antenna measurements and barely vary with frequency. A similar behavior is observed for $\tau_{abs}$, thus confirming our assumption that in the measurements with ports the microwave waveguide system is more opened than in those with antennas. Nevertheless, the strength of \Ti-invariance violation is similar for both measurement procedures. This is expected because the experimental setups are identical except of the procedure of opening the resonator, that is, the number of ferrites, their positions and the size of the external magnetic fields are the same.
We find good agreement between the experimental curves for the autocorrelation functions and distributions of the experimentally measured scattering-matrix elements and the RMT prediction, both for the antenna and the port measurements. Therefore, we show in~\reffig{fig:s12disport} only results obtained from the port measurements. The values of the input parameters, $T_1,T_2,\tau_{abs}$ and $\xi$, are indicated in the panels showing distributions of the scattering matrix elements. The same values were used in the analysis of the RMT results for the autocorrelation functions, shown in~\reffig{fig:corrport}. We may conclude, that the waveguide graph, when considered as an open system exhibits the properties of a typical quantum-chaotic scattering system without or with violated \T invariance. 

\section{CONCLUSION\label{Concl}}
We present experimental results for a waveguide network constructed from waveguides of incommensurate lengths, that are joined at junctions with a relative angle of $120^\circ$. In part of the experiments partial \Ti-invariance violation was induced. We analyzed spectral properties in the frequency range where the analogy to the corresponding quantum waveguide graph holds. In distinction to quantum graphs with Neumann boundary conditions at the vertices and to their experimental realization, namely microwave networks, the vertex scattering matrix describing the transport properties through the vertices or junctions from one bond or waveguide to another one depends on the microwave frequency. It needs to be derived from the wave-function properties at the junctions. A drawback of microwave networks and quantum graphs with Neumann boundary conditions is the occurrence of backscattering at the vertices which also takes place at the waveguide walls in the junctions. The design of waveguide graph, that is, the relative angle of $120^\circ$ was developed in 2014-2015~\cite{Dietz2022} such that the backscattering and frequency dependence are minimized~\cite{Bittner2013}. The waveguide graph is constructed from metal plates with an extension of about 1~m$^2$ and the surface was covered with high-quality copper, so that high quality factors of up to $Q\simeq 6500$ were achieved, such that complete sequences of eigenfrequencies could be identified. 

We come to the conclusion that the spectral properties coincide well with those of the corresponding quantum graph. Backscattering is also present in them, however not as pronounced as in quantum graphs with Neumann boundary conditions. We also investigated statistical properties of the wave functions and confirm the prediction made in~\cite{Kaplan2001} that deviations from RMT predictions can be attributed to wave functions that are localized on a part of the whole graph. Yet, the strength distribution, which only incorporates wave-function components at the positions of the antennas, agrees well with the RMT prediction for quantum systems with a classically chaotic counterpart. The fluctuation properties of the scattering matrix describing the measurement process of the resonance spectra are also well described by those predicted for typical quantum-chaotic scattering systems, both for the cases of preserved and partially violated \Ti-invariance violation. 

These findings suggest, that microwave waveguide graphs may serve as a suitable model for the investigation of features of quantum chaos. Compared to microwave networks, they have the advantage that the boundary conditions at the junctions may be varied, by changing the relative angle of the waveguides at the junctions or by changing material in a controlled way. Furthermore, while in a waveguide graph the electric field, i.e. wave function intensity can be measured in the whole system, this is only possible at the vertices for a microwave network. However, this part of the graph is the most interesting one, because in the waveguides themselves the wave propagation is sinusoidal.

As mentioned in the introduction, a new waveguide system, constructed from a microwave photonic crystal with square lattice structure, was studied numerically with COMSOL Multiphysics~\cite{Ma2021}. A disadvantage of such systems is that the number of modes that exist in the band gap is limited and smaller than that of single transversal modes in a waveguide graph of corresponding geometry.

The range relevant for the simulation of a quantum graph coincides with that of a single transversal mode. We also performed experiments in the frequency range above the cutoff frequency for a second transversal mode. This region will be further explored with focus on a recent work by S. Gutzmann and U. Smilansky~\cite{Gnutzmann2022}. 

\section{Acknowledgement}
This work was supported by the NSF of China under Grant Nos. 11775100, 12047501 and 11961131009. WZ acknowledges financial support from the China Scholarship Council (No. CSC-202106180044). BD and WZ acknowledge financial support from the Institute for Basic Science in Korea through the project IBS-R024-D1.
\bibliography{References}
\end{document}